\documentclass[prc,twocolumn,showpacs,superscriptaddress,preprintnumbers,amsmath,amssymb,nofootinbib]{revtex4-1}
\usepackage{bm}
\usepackage{dcolumn}
\usepackage{multirow}
\usepackage{ifpdf}
\usepackage{url}
\usepackage[utf8]{inputenc}

\ifpdf
  \usepackage{graphicx}     
  \usepackage{hyperref}
\else     
  \usepackage[dvipdfmx]{graphicx}     
  \usepackage[dvipdfmx]{hyperref}
\fi

\newcommand{\thatis}{{\it i.e.}}
\newcommand{\dlangle}{\langle\kern-1.0pt\langle}
\newcommand{\drangle}{\rangle\kern-1.0pt\rangle}

\usepackage[normalem]{ulem}  
\usepackage[dvipdfmx]{color} 

\renewcommand\sout{\bgroup \color{red} \ULdepth=-.5ex \ULset}

\hypersetup{
colorlinks=true,
linkcolor=blue,
citecolor=blue,
urlcolor=blue
}

\begin{document}
\title{Rapidity decorrelation of anisotropic flow caused by hydrodynamic fluctuations}

\author{Azumi Sakai}
\email{a-sakai-s4d@eagle.sophia.ac.jp}
\affiliation{%
Department of Physics, Sophia University, Tokyo 102-8554, Japan
}
\author{Koichi Murase}
\email{murase@pku.edu.cn}
\affiliation{%
Center for High Energy Physics, Peking University, Beijing 100871, China
}
\affiliation{%
Department of Physics, Sophia University, Tokyo 102-8554, Japan
}

\author{Tetsufumi Hirano}
\email{hirano@sophia.ac.jp}
\affiliation{%
Department of Physics, Sophia University, Tokyo 102-8554, Japan
}

\date{\today}

\begin{abstract}
We investigate the effect of hydrodynamic fluctuations
on the rapidity decorrelations of anisotropic flow in high-energy nuclear collisions
using a (3+1)-dimensional integrated dynamical model.
The integrated dynamical model consists of twisted initial conditions,
fluctuating hydrodynamics, and hadronic cascades on an event-by-event basis.
To understand the rapidity decorrelation,
we analyze the factorization ratio in the longitudinal direction.
Comparing the factorization ratios
between fluctuating hydrodynamics and ordinary viscous hydrodynamics,
we find a sizable effect of hydrodynamic fluctuations on rapidity decorrelations.
We also propose to calculate the Legendre coefficients of the flow magnitude and the event-plane angle
to understand the decorrelation of anisotropic flow in the longitudinal direction.
\end{abstract}


\maketitle

\section{Introduction}
High-energy nuclear collision experiments have been performed
at Relativistic Heavy Ion Collider (RHIC) at Brookhaven National Laboratory
and at Large Hadron Collider (LHC) at CERN
to understand bulk and transport properties of the deconfined nuclear matter, the quark gluon plasma (QGP)~\cite{Yagi:2005yb}.
A vast body of the experimental data has been taken at RHIC and LHC\@.
Among them, the large magnitude of the second-order azimuthal anisotropy of the
emitted hadrons, also known as the elliptic flow parameter \cite{Ollitrault:1992bk},
is one of the major discoveries
at RHIC~\cite{Ackermann:2000tr,Adler:2001nb,Adams:2003am,Adcox:2002ms,Adler:2003kt,Back:2002gz}.
These data were consistent with predictions and/or postdiction
from ideal hydrodynamic models~\cite{Kolb:2000fha,Huovinen:2001cy,Teaney:2000cw,Teaney:2001av,Hirano:2001eu,Hirano:2002ds},
which leads to the discovery of the almost perfect fluidity of the QGP fluids~\cite{Heinz:2001xi,Lee:2005gw,Gyulassy:2004zy,Shuryak:2004cy,Hirano:2005wx}.
The large elliptic flow parameters were also confirmed
even at higher collision energies at LHC~\cite{Aamodt:2010pa,Chatrchyan:2011pb,ATLAS:2011ah}.
The higher-order anisotropic flow parameters have also been measured
at RHIC~\cite{Adare:2011tg,Adam:2019woz}
and LHC~\cite{ALICE:2011ab,ATLAS:2012at,Chatrchyan:2012wg}.

These anisotropic flow parameters are useful measures to determine the properties of the created QGP fluids.
They are sensitive not only to the collision geometry~\cite{Ollitrault:1992bk}
and the event-by-event fluctuations~\cite{Alver:2010gr} of the initial transverse profiles
but also to viscosities of the QGP fluids.
So far, relativistic hydrodynamic models with viscosities have made
a huge success in understanding
these anisotropic flow parameters~\cite{Song:2007ux, Dusling:2007gi, Luzem:2008cw, Schenke:2010nt, Bozek:2010wt, Song:2010mg, Song:2011qa, Schenke:2011tv}.
Active studies are going on to extract the transport properties of the QGP
such as the shear and bulk viscosity coefficients.
Recently Bayesian analysis technique has been employed to constrain
the various model parameters such as transport coefficients and initial state parameters
within current viscous fluid dynamical models~\cite{Bernhard:2016tnd,Bernhard:2019bmu}.
In this direction, more systematic studies would be needed
for rigorous determination of these physical parameters
by sophisticating the dynamical models
to include important physics such as longitudinal dynamics and thermal fluctuations.

The thermal fluctuations of the systems described by hydrodynamics are called hydrodynamic fluctuations.
In the framework of hydrodynamics, microscopic degrees of freedom are integrated out by coarse-graining,
and only a few macroscopic variables, such as flow velocities and thermodynamic fields, remain as slow dynamical variables.
However, the macroscopic dynamics cannot be completely separated from the microscopic one
when the scale of interest is close to the microscopic scale.
The microscopic dynamics induces the fluctuations of the macroscopic variables around its coarse-grained values
on an event-by-event basis, which is nothing but the thermal fluctuations.
Because of the insufficient scale separation
of the relevant macroscopic hydrodynamics of the created matter and the microscopic dynamics in the collision processes,
hydrodynamic fluctuations would be of significant importance in the dynamical models of the high-energy nuclear collisions.
So far, almost all the hydrodynamic models applied to the analysis of
experimental data are based on conventional viscous hydrodynamics without hydrodynamic fluctuations.
Towards a thermodynamically consistent description of the event-by-event dynamics of thermodynamic fields,
it is important to develop dynamical models for the high-energy nuclear collisions
based on relativistic fluctuating hydrodynamics~%
\cite{Calzetta:1997aj,Kapusta:2011gt,Murase:2013tma,HF,muraseD,Murase:2019cwc,
Akamatsu:2016llw,Akamatsu:2017rdu,Martinez:2018wia,An:2019osr},
which is the viscous hydrodynamics with hydrodynamic fluctuations.
The fluctuation--dissipation relation (FDR) tells us
the magnitude of the thermal fluctuations is determined by the corresponding dissipation rates.
Therefore the system maintains its proper thermodynamic distributions near the equilibrium
by the balance of effects of the dissipations and fluctuations.
Since fluctuations cause the deviation of thermodynamic variables
while the dissipation pulls it back to its mean value on an event-by-event basis,
fluctuations as a whole broaden the distributions of the thermodynamic variables
while the dissipation narrows the distribution.
In the conventional viscous fluid dynamics, dissipative currents such as the shear stress tensor
and the bulk pressure are driven by thermodynamic forces. In addition, they fluctuate around
these systematic forces in fluctuating hydrodynamics,
and the power of these fluctuating forces is given by the FDR\@.

In this paper,
we focus on how the hydrodynamic fluctuations affect
the longitudinal dynamics of the QGP fluids produced in high-energy nuclear collisions.
The initial profile of the QGP fluids possesses a strong correlation
in the direction of the collision axis (the longitudinal direction),
which comes from the color flux tubes or hadron strings
stretched in the longitudinal direction between the two collided nuclei.
This longitudinal correlation in the initial stage can be observed in
approximately boost-invariant transverse profiles and
participant-plane angles aligned for different rapidities in each event.
Since the QGP fluids respond to anisotropy of the initial transverse profiles in expansion,
the resultant event-plane angle of the
azimuthal momentum distribution in the final stage
is expected to be almost independent of (pseudo-)rapidity.
However, when the hydrodynamic fluctuations are taken into account,
this naive picture should be modified.
Since the hydrodynamic fluctuations arise at each space-time point independently
without any longitudinal correlations,
they disturb the spatial longitudinal structure of the initial stage.
This enhances the fluctuation of the event-plane angles
of final azimuthal anisotropies at different rapidities.
Thus the hydrodynamic fluctuations are expected to
decrease the longitudinal correlation.

To investigate this rapidity decorrelation phenomenon
caused by the hydrodynamic fluctuations,
we consider the observable called factorization ratios
which quantify the decorrelation.
The factorization ratio was initially proposed as a function of transverse momentum~\cite{Gardim:2012im}
and was later extended in the longitudinal direction
as a function of pseudorapidity~\cite{Khachatryan:2015oea}
for the purpose of analyzing the longitudinal dynamics.
Since then, the factorization ratios in the longitudinal direction
are widely measured in experiments~\cite{Khachatryan:2015oea, Aaboud:2017tql, Huo:2017hjv,
Nie:2019bgd, Aad:2020gfz}.
The origin of the rapidity decorrelation is understood from decorrelation of
event-plane angle and/or magnitude of flow.
To separate the effects of these two decorrelations,
the factorization ratios are defined in different ways~\cite{Aaboud:2017tql,Bozek:2017qir}.
These factorization ratios are studied in various models with the effects of
initial twists~\cite{Bozek:2017qir, Bozek:2010vz},
longitudinal fluctuations~\cite{Wu:2018cpc, Jia:2017kdq, Bozek:2015bna, Pang:2014pxa, Pang:2015zrq, Pang:2018zzo},
glasma~\cite{Schenke:2016ksl},
and dynamical initial states~\cite{Shen:2017bsr}.
The factorization ratios can be reasonably described by the effects of initial twists~\cite{Bozek:2017qir}.
Also, the factorization ratios are affected by the length of the initial string structures~\cite{Wu:2018cpc}
which depends on the collision energy.
The effects of eccentricity decorrelation in various collision systems are investigated recently~\cite{Behera:2020mol}.
Although these models exhibit the factorization breakdown,
they have not yet quantitatively described all the measurements,
including all the centrality, different harmonic orders, and the collision energy dependence,
in a single model.
So far, the effects of the hydrodynamic fluctuations on the factorization ratio have not been studied systematically.
In this paper, we investigate the effects of this mechanism of the rapidity decorrelation by the hydrodynamic fluctuations.
We also propose to calculate the Legendre coefficients of the flow magnitude and the event-plane angle of
anisotropic flow parameters as functions of pseudorapidity to characterize
the event-by-event longitudinal structure of the QGP fluid evolution.

This paper is organized as follows:
We first review in Sec.~\ref{sec:model}
the integrated dynamical model highlighting the framework of relativistic fluctuating hydrodynamics.
In Sec.~\ref{sec:analysis and results}, after describing the details of the model parameter tuning,
we show results of factorization ratios in the longitudinal direction.
We also propose and calculate the Legendre coefficients of anisotropic flow parameters.
Finally Sec.~\ref{sec:summary} is devoted to the summary of the present study.
We use the natural units, $\hbar = c = k_{B} = 1$, and the Minkowski metric,
$g_{\mu\nu} = \mathrm{diag}(1, -1, -1, -1)$,
throughout this paper.

\section{Model}
\label{sec:model}
In this paper, we employ the integrated dynamical model discussed in Ref.~\cite{ppnp}
to describe space-time evolution of high-energy nuclear collisions.
The integrated dynamical model in the present study
is composed of three models corresponding to three stages of collision reactions:
a Monte-Carlo version of the Glauber model extended in the longitudinal direction~\cite{Hirano:2005xf}
for the entropy production in the initial stage
which is implemented in the code {\tt mckln},
relativistic fluctuating hydrodynamic model {\tt rfh}~\cite{muraseD} for space-time evolution of matter close to equilibrium
which is dissipative hydrodynamics with causal hydrodynamic fluctuations,
and a hadron cascade model {\tt JAM}~\cite{Nara:1999dz} for a microscopic transport of hadron gases.

\subsection{Causal fluctuating hydrodynamics}

To describe the space-time evolution of fluids in the intermediate stage of collisions,
we use the relativistic fluctuating hydrodynamics model {\tt rfh}~\cite{muraseD}
which solves the second-order causal fluctuating hydrodynamic equations
in which the hydrodynamic fluctuations are treated as stochastic terms.
We consider three types of hydrodynamics to quantify
the effects of the hydrodynamic fluctuations and dissipations.
In \textit{fluctuating hydrodynamic models},
we include both the shear viscosity and the corresponding hydrodynamic fluctuations
with several choices of the cutoff parameter.
For comparison, we also run the \textit{viscous hydrodynamic model}
in which the fluctuations are turned off,
and also the \textit{ideal hydrodynamic model}
in which both the fluctuations and the dissipations are turned off.

The main dynamical equations of the second-order fluctuating hydrodynamics
are the same with the usual hydrodynamics,
the conservation law of energy and momentum:
\begin{align}
  \partial_\mu T^{\mu\nu} &= 0,
\end{align}
where $T^{\mu\nu}$ is the energy--momentum tensor of fluids,
which is written in terms of thermodynamic variables as
\begin{align}
  T^{\mu \nu} &= (e+P)u^\mu u^\nu - P g^{\mu \nu} + \pi^{\mu \nu}.
\label{eq:tensor_decom_Tmunu}
\end{align}
Here $e$ is the energy density, $P$ is the pressure, and
$\pi^{\mu \nu}$ is the shear stress tensor.
It should be noted here that the stochastic terms are included
in the shear stress $\pi^{\mu\nu}$ in this formalism for the second-order theory.
We employ the Landau frame to define the fluid velocity $u^\mu$ as
$T^\mu_{\ \nu} u^\nu = e u^\mu$
so that no energy current appears in Eq.~\eqref{eq:tensor_decom_Tmunu}.
We do not consider the bulk pressure in the present study.
We also neglect the conservation law of the baryon number
since we focus on high-energy nuclear collisions
at the LHC energies at which the baryon number is
expected to be small around midrapidity.
For an equation of state,
we employ
\textit{s}95\textit{p}-v1.1~\cite{Huovinen:2009yb},
which is a smooth interpolation
of the equation of state from (2+1)-flavor lattice QCD simulations
and that from the hadron resonance gas
corresponding to the hadronic cascade model described in Sec.~\ref{sec:hadron-cascade}.

In the second-order fluctuating hydrodynamics,
the fluctuation terms can be introduced
in the constitutive equations for the shear-stress tensor~\cite{muraseD,Murase:2019cwc}:
\begin{multline}
  \label{eq:shear_stress_tensor}
  \tau_\pi{\Delta^{\mu\nu}}_{\alpha\beta}u^{\lambda}\partial_{\lambda}\pi^{\alpha\beta}
  + \left(1+\frac{4}{3}\tau_\pi\partial_{\lambda}u^\lambda \right)\pi^{\mu\nu} \\
  = 2\eta{\Delta^{\mu\nu}}_{\alpha\beta}\partial^\alpha u^\beta+\xi^{\mu\nu},
\end{multline}
where $\eta$ is the shear viscosity, $\tau_{\pi}$ is the relaxation time,
and the tensor ${\Delta^{\mu\nu}}_{\alpha\beta}=\frac{1}{2}\left({\Delta^\mu}_\alpha{\Delta^\nu}_\beta+{\Delta^\mu}_\beta{\Delta^\nu}_\alpha \right)-\frac{1}{3}{\Delta^{\mu\nu}}\Delta_{\alpha\beta}$
is a projector for second-rank tensors onto the symmetric and traceless components
transverse to the flow velocity.
The noise term $\xi^{\mu \nu}$, which represents the hydrodynamic fluctuations,
is given as random fields of the Gaussian distribution in which autocorrelation is determined
by the fluctuation--dissipation relation,\footnote{
  The second-order modification terms of the fluctuation--dissipation relation
  in arbitrary backgrounds derived in Ref.~\cite{Murase:2019cwc}
  are not included in the present study.}
\begin{align}
  \label{eq:FD}
  \langle\xi^{\mu\nu}(x)\xi^{\alpha\beta}(x')\rangle = 4\eta(x) T(x)\Delta^{\mu\nu\alpha\beta}
 \delta^{(4)}(x-x').
\end{align}
Here the angle brackets mean ensemble average
and $T$ is the temperature.
Note that the ensemble average of the noise term vanishes by definition,
\begin{align}
  \langle \xi^{\mu\nu}(x) \rangle  = 0.
\end{align}

In the hydrodynamic simulations,
we employ the Milne coordinates $(\tau, \eta_s, x, y)\equiv(\tau, \eta_s, \bm{x}_\perp)$,
where $\tau = \sqrt{t^2-z^2}$ is the proper time  and
$\eta_s = \tanh^{-1}(z/t)$ is the space-time rapidity.
Here $z$ is a coordinate along the collision axis
and we assume the Lorentz-contracted two nuclei collide with each other at
$z=0\ \text{fm}$ and $t=0\ \text{fm}$.  The origin in the transverse plane
is taken to be the center of mass of the participant nucleons
with $x$-axis parallel to the impact parameter vector.

In the Milne coordinates, the fluctuation--dissipation relation \eqref{eq:FD} is rewritten as
\begin{multline}
  \label{eq:FD_Milne}
  \langle\xi^{\mu\nu}(\tau, \eta_s, \bm{x}_\perp)\xi^{\alpha\beta}(\tau', \eta_s', \bm{x}'_\perp)\rangle \\
  = 4\eta T\Delta^{\mu\nu\alpha\beta}
  \frac1{\tau}\delta(\tau-\tau')\delta(\eta_s-\eta_s')\delta^{(2)}(\bm{x}_\perp - \bm{x}_\perp'),
\end{multline}
where the factor $1/\tau$ comes from the Jacobian of
variable transformation from the Cartesian to the Milne coordinates.
Here, the Lorentz indices, such as $\mu$ and $\alpha$, run over $\tau$, $\eta_{s}$, $x$, and $y$.
In the actual calculations, to avoid the ultraviolet singularity,
we generate a smeared fluctuation term
by considering the convolution with the Gaussian kernel:
\begin{multline}
\label{eq:Gaussian_with_cutoff}
  G(\eta_s, \bm{x}_\perp; \lambda_\eta,\lambda_\perp) \\
  = \frac1{\sqrt{2\pi\lambda_\eta^2}}\frac1{2\pi\lambda_\perp^2}
    \exp\left(- \frac{\eta_s^2}{2\lambda_\eta^2}
    - \frac{\bm{x}_\perp^2}{2\lambda_\perp^2}\right),
\end{multline}
where $\lambda_\eta$ and $\lambda_\perp$ are
the cutoff parameters in the longitudinal and transverse directions, respectively.
The detailed procedure of smeared noise generation is described in Ref.~\cite{muraseD}.
It should be noted here that these cutoff parameters
effectively control the magnitude of the fluctuations:
The smaller cutoff length is taken, the larger magnitude of fluctuations becomes.
Note also that the smearing in the temporal direction is not introduced in this study.

\subsection{Initial condition}
\label{sec:initial_condition}
For the initial conditions of the causal fluctuating hydrodynamics,
we parametrize the entropy density distributions, $s(\tau_0, \eta_s, \bm{x}_\perp)$,
at the hydrodynamic initial proper time $\tau_0$.
For the parametrization of the entropy distribution,
we employ the Monte--Carlo version of the Glauber model
in the transverse plane~\cite{GLAUBER20063,ppnp}
and combine it with the modified Brodsky--Gunion--Kuhn (BGK) model
in the longitudinal direction~\cite{Brodsky:1977de,Hirano:2005xf}.
Using the Monte--Carlo Glauber model,
we calculate the transverse profiles
of the participant number densities of nuclei A and B,
$\rho_{\mathrm{part}}^\mathrm{A}(\bm{x}_\perp)$ and
$\rho_{\mathrm{part}}^\mathrm{B}(\bm{x}_\perp)$, respectively,
and the number density of the binary collisions,
$\rho_{\mathrm{coll}}(\bm{x}_\perp)$, for a randomly sampled impact parameter $b$.
In the modified BGK model,
the idea of ``rapidity triangle'' or
``rapidity trapezoid''~\cite{Brodsky:1977de, Adil:2005qn, Hirano:2005xf}
is demonstrated by parametrizing the space-time rapidity dependent
entropy density distribution as
\begin{align}
  &s(\tau_0, \eta_s, \bm{x}_\perp)
   = \frac{C}{\tau_0}\theta(Y_b-\left| \eta_s \right| ) f^{pp}(\eta_s)
    \biggl[
 \alpha\rho_{\mathrm{coll}}(\bm{x}_\perp) \nonumber \\
  &     + \frac{1-\alpha}{2}\left(
\frac{Y_b\!-\!\eta_s}{Y_b}\rho_{\mathrm{part}}^{\mathrm{A}}(\bm{x}_\perp)
    +  \frac{Y_b\!+\!\eta_s}{Y_b}\rho_{\mathrm{part}}^{\mathrm{B}}(\bm{x}_\perp) \right)
\biggr],
  \label{eq:N}
\end{align}
where parameters
$Y_{b}$, $C$, and $\alpha$ are the beam rapidity, the normalization factor,
and the hard fraction, respectively.
The function $\theta(x)$ is the Heaviside step function to cut off the profile beyond
the beam rapidity.
The function $f^{pp}(\eta_s)$ models a longitudinal profile of proton--proton collisions,
for which we use the following form:
\begin{align}
  f^{pp}(\eta_s)=\exp\left[ -\theta(|\eta_s|-\Delta\eta)\frac{(|\eta_s|-\Delta\eta)^2}{\sigma^2_\eta}\right],
  \label{eq:sigmadelta}
\end{align}
where $\Delta\eta$ and $\sigma_\eta$ are model parameters.
The function $f^{pp}(\eta_s) = 1$ when $|\eta_s|<\Delta\eta$, and
decreases with the Gaussian when $|\eta_s|\ge\Delta\eta$.
The parameter $\sigma_\eta$ controls the Gaussian width.
It would be instructive to see the initial entropy
at midrapidity $\eta_{s}=0$ in the coordinate space scales with a linear
combination of the number of participants $N_\mathrm{part}$
and the number of binary collisions $N_\mathrm{coll}$ as
\begin{align}
\frac{dS}{d\eta_{s}}(\tau_0, \eta_{s}=0)
& = \int d^2\bm{x}_\perp \tau_{0} s(\tau_0, \eta_{s}=0, \bm{x}_\perp) \nonumber \\
& = C \left[ \alpha N_{\mathrm{coll}} + (1-\alpha)\frac{N_{\mathrm{part}}}{2} \right].
\end{align}
We determine the model parameters $C$, $\alpha$, $\Delta\eta$, and $\sigma_\eta$
comparing the results with experimental data
of the centrality and the pseudorapidity dependences of charged-hadron multiplicity (see Sec.~\ref{sec:model parameter}).

In the modified BGK model,
the initial entropy density is asymmetric
with respect to space-time rapidity
when the number densities of participants  at
the same transverse position $\bm{x}_\perp$ are different between nuclei A and B, \thatis,
$\rho_{\mathrm{part}}^{\mathrm{A}}(\bm{x}_\perp) \not= \rho_{\mathrm{part}}^{\mathrm{B}}(\bm{x}_\perp)$.
This asymmetry brings a twist structure
to the entropy density distribution
in the reaction plane ($x$-$\eta_s$ plane)~\cite{Adil:2005qn}.
This twist structure results in a rapidity-dependent eccentricity~\cite{Hirano:2001eu}
and participant-plane angle.
The rapidity dependence of eccentricity and participant-plane angle
is important in understanding the rapidity decorrelation of anisotropic flow
as we will discuss in Sec.~\ref{sec:Factorization ratio}.

For the initial flow velocity, we put the Bjorken scaling solution~\cite{Bjorken:1982qr},
$u^\mu(\tau_0) = (\cosh \eta_{s}, 0, 0, \sinh \eta_{s})$ in the Cartesian coordinates.
This means there are no fluctuations in flow velocity at $\tau=\tau_0$.
It should be also noted that there are no longitudinal fluctuations of the initial profiles in this model
since Eq.~\eqref{eq:sigmadelta} is a smooth function of space-time rapidity $\eta_{s}$.
We set the initial shear stress tensor to
$\pi^{\mu\nu}(\tau_0) = 0$ in the present study.


\subsection{Particlization and hadron cascade}
\label{sec:hadron-cascade}
After macroscopic hydrodynamic simulations, we switch the description to the microscopic kinetic theory
for hadrons at a switching temperature, $T_{\mathrm{sw}}$.
For the space-time evolution of hadron gases, we employ a microscopic transport model, {\tt JAM}~\cite{Nara:1999dz}.
This model deals with hadronic rescattering and decay in the late stage of collisions.

For the initialization of hadrons in {\tt JAM},
we sample hadrons from fluid elements in the hypersurface determined by $T(x) = T_{\mathrm{sw}}$.
We specify the four-momentum, $p^\mu=(E, \bm{p})$, and position, $x^\mu=(t, \bm{x})$,
using the Cooper--Frye formula~\cite{Cooper:1974mv} with a viscous correction~\cite{Teaney:2003kp,Monnai:2009ad},
\begin{align}
\label{eq:Cooper-Frye}
\Delta N_i &= g_{i} \int \frac{d^3\bm{p}}{(2 \pi)^3 E} p \cdot \Delta\sigma f_{i}(p, x)\theta(f_{i}(p, x)),\\
f_{i}(p, x) &= f_{0, i}(p,x) + \delta f_{i}(p, x),\\
\label{eq:viscorrection}
f_{0, i}(p,x) &= \frac{1}{\exp(p\cdot u/T_{\mathrm{sw}}) - \epsilon},\\
\delta f_{i}(p, x) &= f_{0, i}(p,x) [1+\epsilon f_{0, i}(p,x)]\frac{\pi^{\mu\nu}p_\mu p_\nu}{2(e+P)T^2}.
\end{align}
Here $g_i$ is the degeneracy for hadron species $i$,
$\Delta\sigma_\mu$ is the switching hypersurface element,
and $\epsilon$ is a factor for fermions ($-1$) or bosons ($+1$).
In this ``particlization'' prescription,
we calculate all hadrons in the list of hadronic cascade code, {\tt JAM}\@.
Since one cannot treat the negative number
of the phase space distribution in the hadron transport model,
we consider the out-going hadrons only
and omit the in-coming hadrons, which bring
the negative number in the Cooper--Frye formula \eqref{eq:Cooper-Frye}.
Simulations of the hadron transport model
are performed until scattering between hadrons and decay of resonances no longer happen.
We switch off the weak and electromagnetic decays,
hence the final hadron distributions are compared directly with experimental data in which
those contributions are corrected.

\section{Analysis and Results}
\label{sec:analysis and results}

Using the integrated dynamical model explained in the previous section, we perform simulations
of Pb+Pb collisions at $\sqrt{s_{NN}}=2.76\ \text{TeV}$.
For each set of the model parameters,
we generate 4000 hydrodynamic events and perform 100 independent particlization and hadronic cascades for each hydrodynamic event.
Thus we obtain $400000 (=4000 \times 100)$ events in total
corresponding to minimum bias events in experiments.
It should be noted that this oversampling prescription reduces computational costs largely.
Analyzing the phase space distributions of final hadrons,
we compare our results with experimental data.
As for the centrality cut, we do almost the same way
as done in the experimental analysis: We categorize
 all the events into each centrality bin
using the charged-hadron multiplicity distribution in $2.9 < |\eta_p| <5.2$~\cite{Khachatryan:2015oea, Chatrchyan:2011pb}.
For each event, we define the centrality percentile to the total number of events.
For example, we regard the top 40000 high-multiplicity events
as the event class at 0--10\% centrality.

In this section, after describing the setup of the model parameters,
we analyze factorization ratios, $r_n(\eta_p^a, \eta_p^b)$, of charged hadrons
and the Legendre coefficients of both the flow magnitude and the event-plane angle as functions of pseudorapidity.

\subsection{Model parameters}
\label{sec:model parameter}
\begin{figure}[htbp]
\begin{center}
\centering
\includegraphics[width=0.5\textwidth,  bb=60 0 320 220]{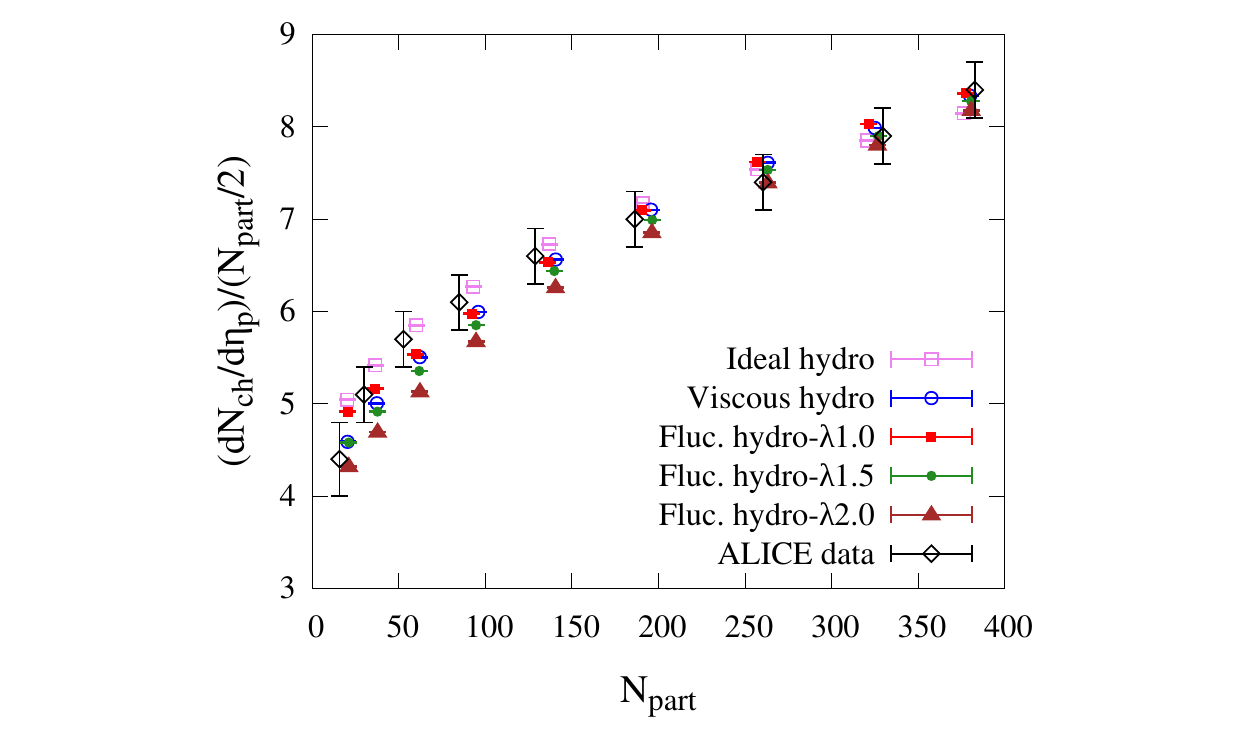}
\caption{(Color Online)
Charged-hadron multiplicity
normalized by the number of the participant pair,
$(dN_\mathrm{ch}/d\eta_p)/(N_\mathrm{part}/2)$,
as a function of the number of participants.
The results from ideal hydrodynamics (open square), viscous hydrodynamics (open circle),
fluctuating hydrodynamics--$\lambda1.0$ (filled square),
fluctuating hydrodynamics--$\lambda1.5$ (filled circle),
and fluctuating hydrodynamics--$\lambda2.0$ (filled triangle) are compared with experimental data (open diamond) obtained by the ALICE Collaboration~\cite{Aamodt:2010cz}.
}
\label{fig:multiplicity}
\end{center}
\end{figure}

\begin{figure*}[htbp]
\begin{center}
\includegraphics[width=0.48\textwidth, bb=30 0 370 220]{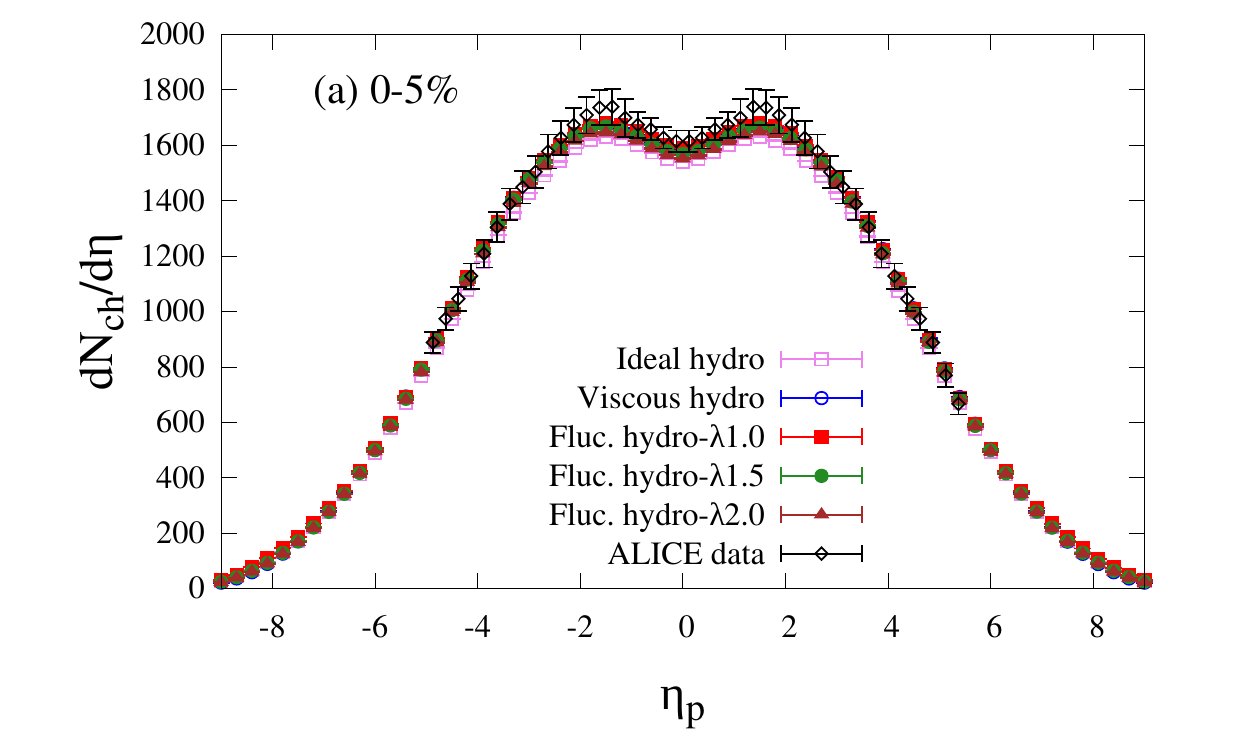}
\includegraphics[width=0.48\textwidth, bb=30 0 370 220]{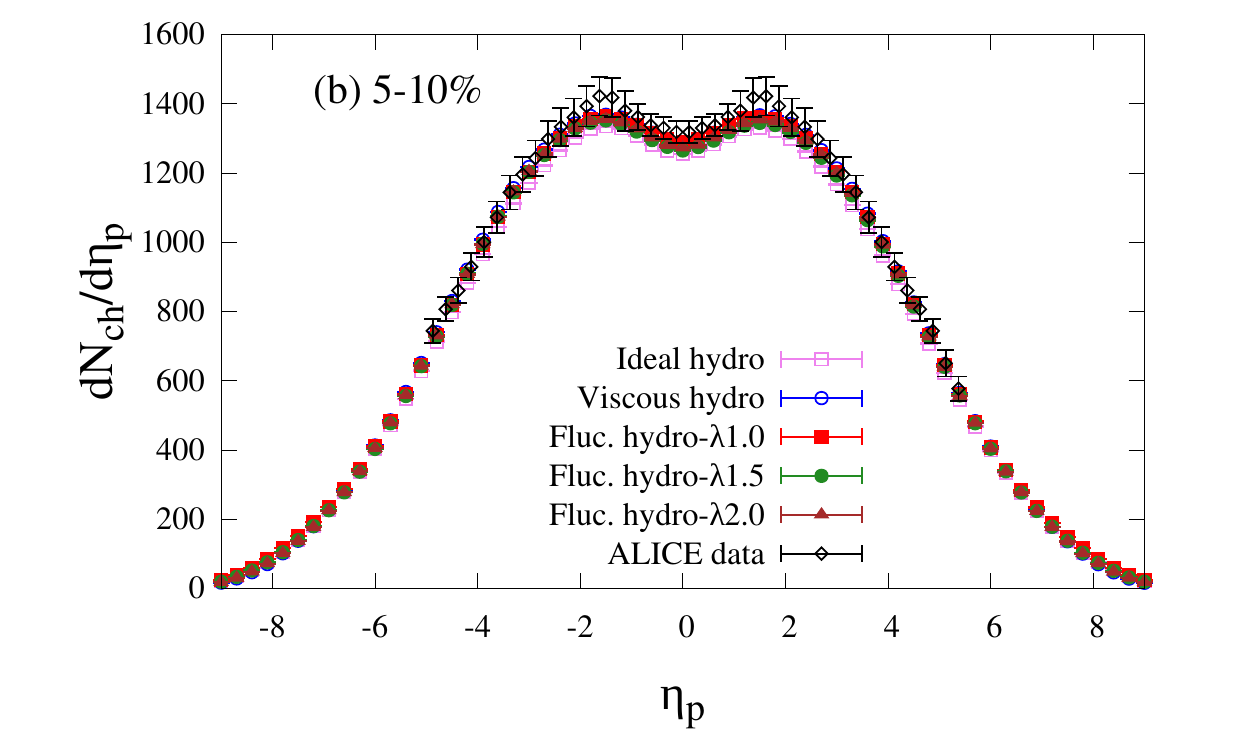}

\includegraphics[width=0.48\textwidth, bb=30 0 370 220]{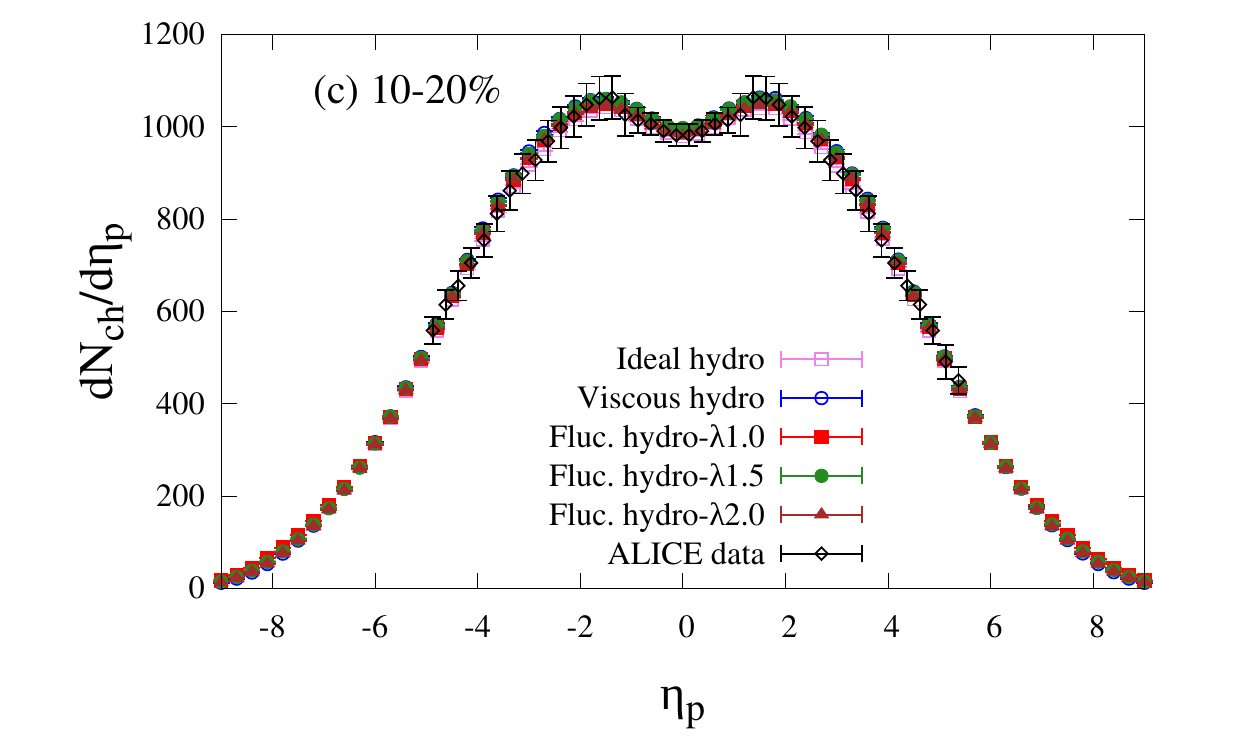}
\includegraphics[width=0.48\textwidth, bb=30 0 370 220]{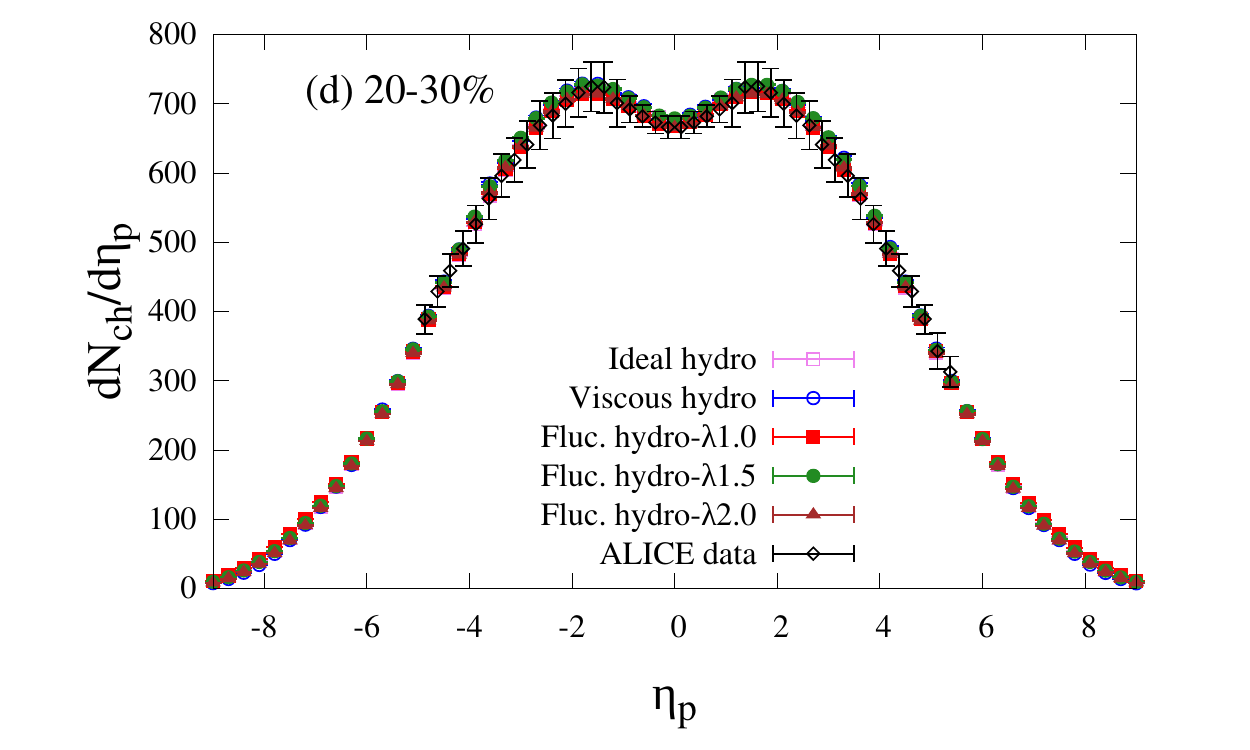}

\includegraphics[width=0.48\textwidth, bb=30 0 370 220]{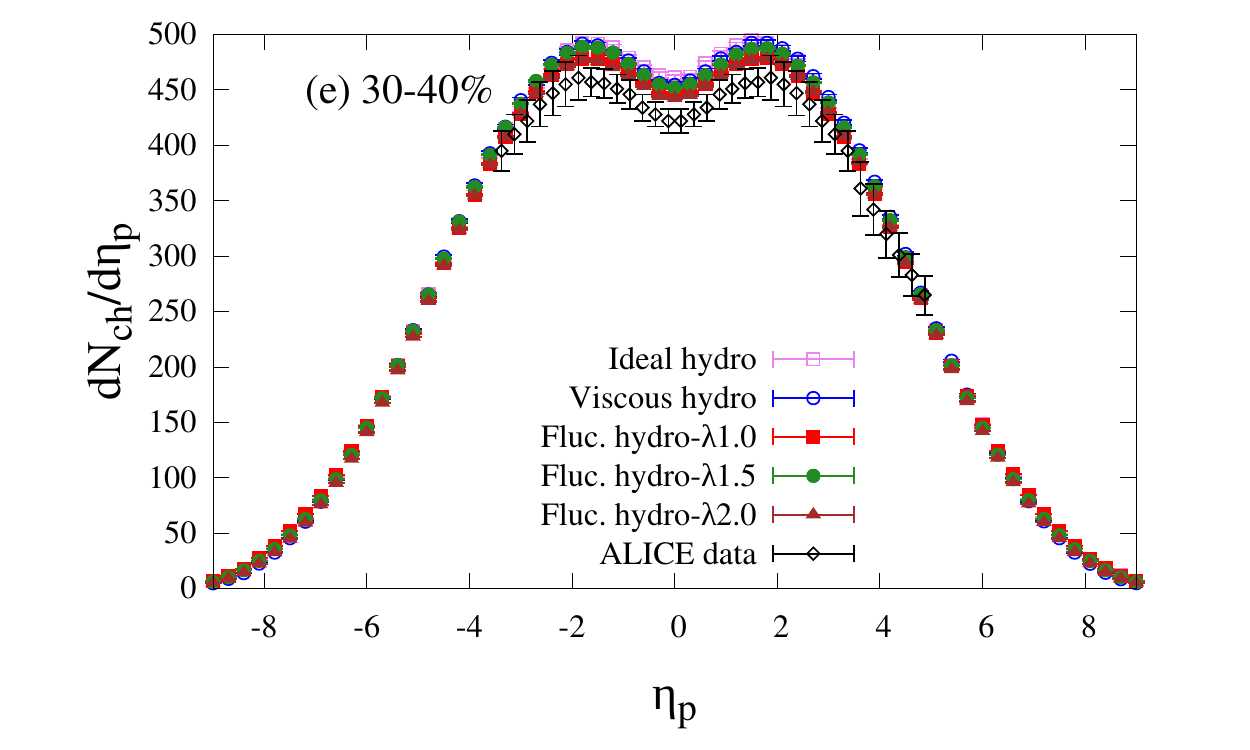}
\includegraphics[width=0.48\textwidth, bb=30 0 370 220]{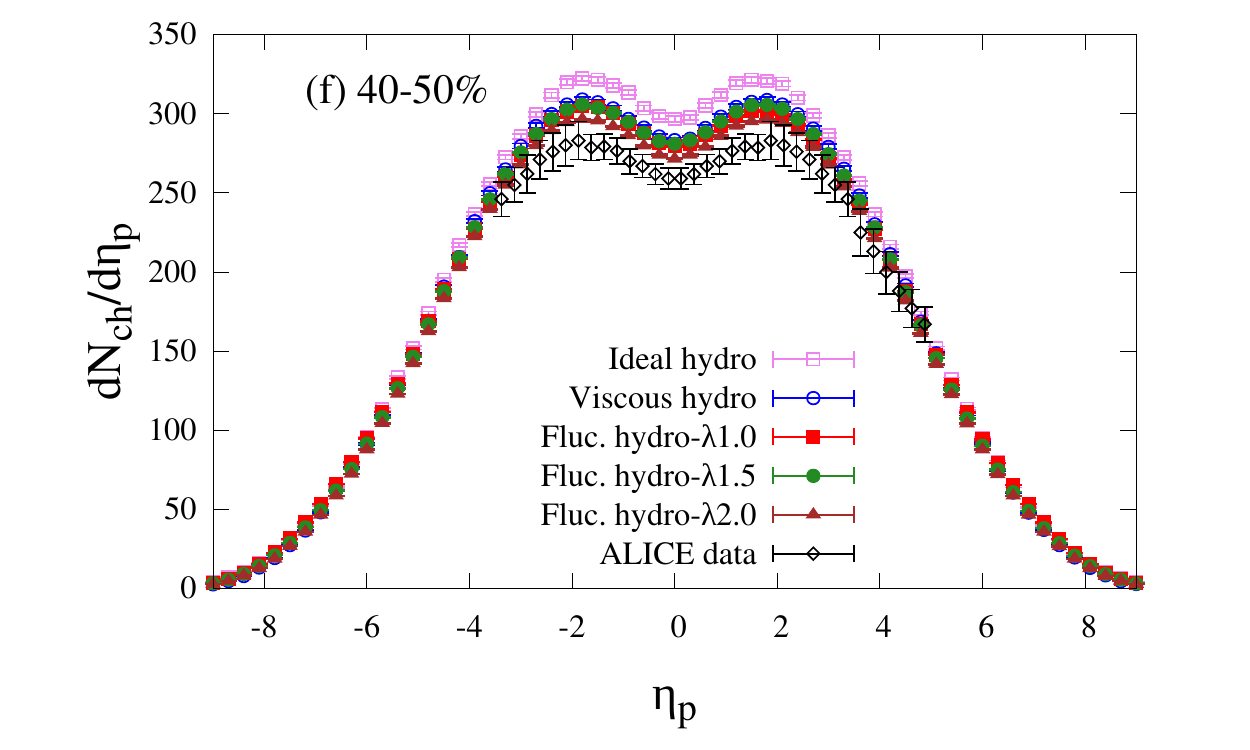}

\includegraphics[width=0.48\textwidth, bb=30 0 370 220]{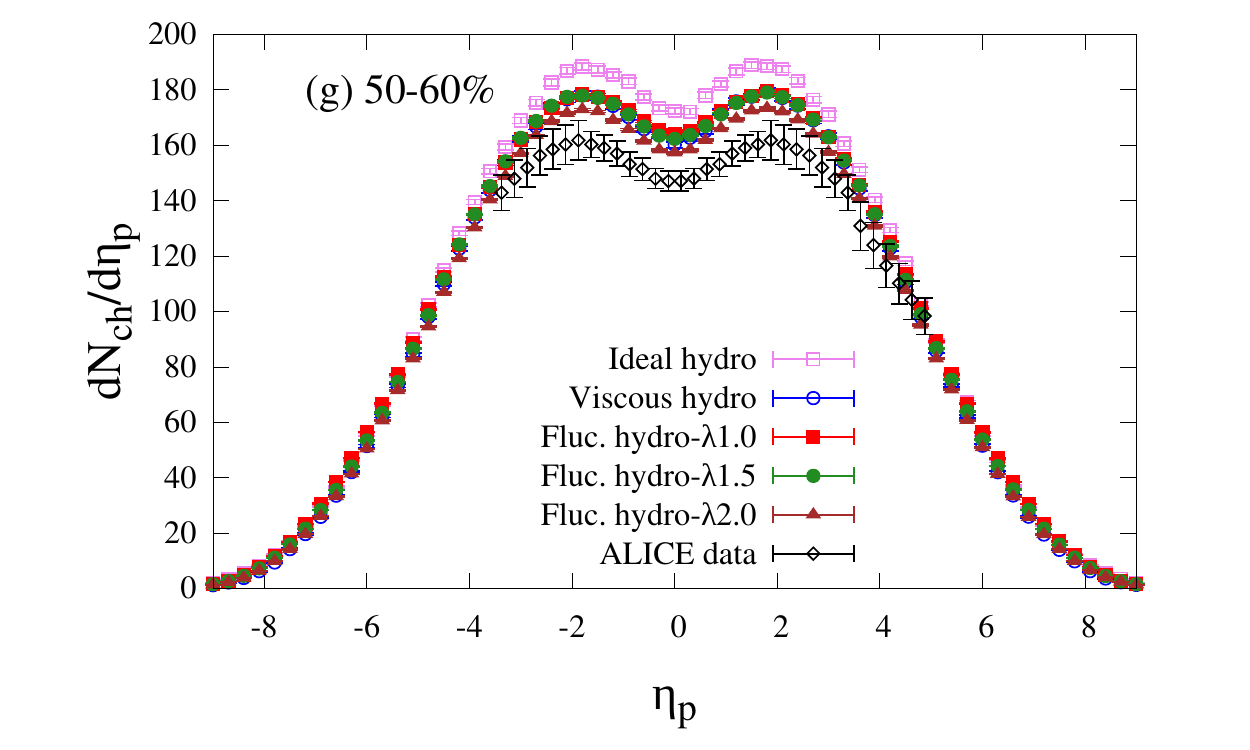}
\includegraphics[width=0.48\textwidth, bb=30 0 370 220]{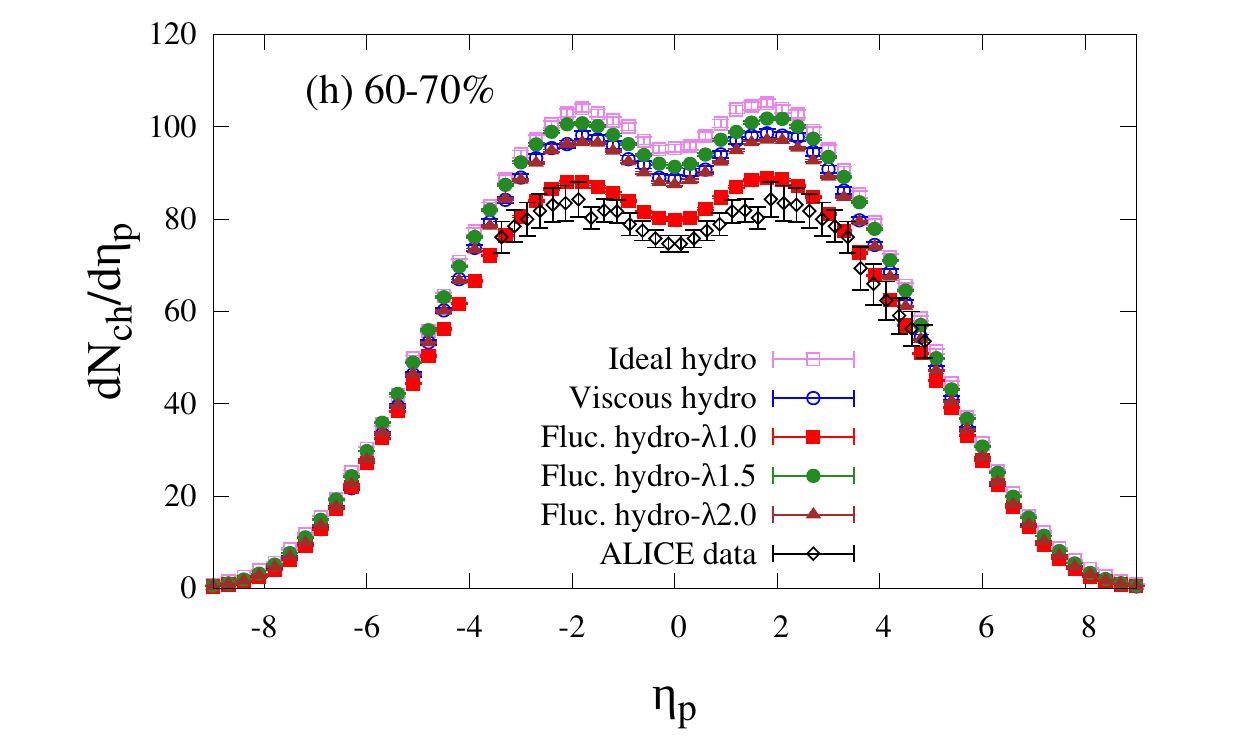}
\caption{(Color Online) Pseudorapidity distributions of charged hadrons
in Pb+Pb collisions at $\sqrt{s_{NN}} = 2.76$ TeV.
Centralities are (a) 0--5\%, (b) 5--10\%, (c) 10--20\%, (d) 20--30\%,
(e) 30--40\%, (f) 40--50\%, (g) 50--60\%, and (h) 60--70\%.
The symbols are the same as in Fig.~\ref{fig:multiplicity}.
The experimental data are taken from Refs.~\cite{Abbas:2013bpa,Adam:2015kda}.
}
\label{fig:multiplicityeta}
\end{center}
\end{figure*}

\begin{figure}[htbp]
\begin{center}
\includegraphics[width=0.5\textwidth,  bb=60 0 320 220]{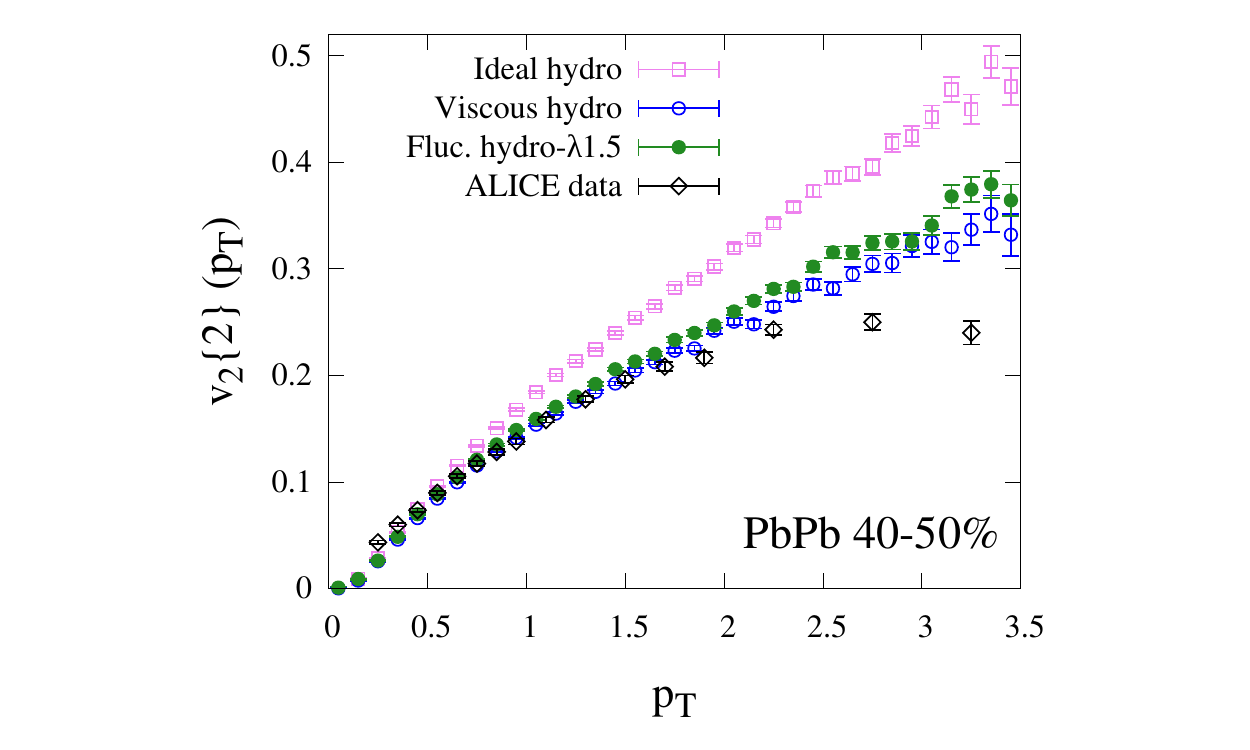}
\caption{(Color Online) $v_{2}\{2\} (p_T)$ in Pb+Pb collisions at $\sqrt{s_{NN}} = 2.76\ \text{TeV}$ for centrality 40--50\%.
The results from ideal hydrodynamics (open square), viscous hydrodynamics (open circle),
and fluctuating hydrodynamics--$\lambda1.5$ (filled circle)
are compared with experimental data (open diamond) obtained by the ALICE Collaboration~\cite{Aamodt:2010pa}.
}
\label{fig:v2}
\end{center}
\end{figure}

In this paper, we set $\eta/s = 1/4\pi$ for shear viscosity~\cite{Kovtun:2004de}
and $\tau_{\pi} = 3/4\pi T$ for relaxation time~\cite{Song:2009gc,Baier:2007ix}.
As we will see in Fig.~\ref{fig:v2}, the shear viscosity is mostly constrained from
experimental data of $p_{T}$-differential elliptic flow parameters in non-central collisions.
We choose the initial proper time $\tau_0 = 0.6\ \text{fm}$ and
the switching temperature $T_{\mathrm{sw}}=155\ \text{MeV}$ as in the previous calculations~\cite{ppnp}.
We tune initial parameters $\Delta\eta$, $\sigma_\eta$, $C$, and $\alpha$
to reproduce the centrality and the pseudorapidity dependences of charged-hadron multiplicity
as we will show in Figs.~\ref{fig:multiplicity} and \ref{fig:multiplicityeta}.
The values of $\sigma_\eta$ and $\Delta\eta$ are tuned as 3.2 and 1.9, respectively, irrespective of hydrodynamic models.
The values of $C$ and $\alpha$ are tuned for each hydrodynamic model
as summarized in Table~\ref{table:parameter}\@.
We run simulations using the ideal hydrodynamic model, the viscous hydrodynamic model,
and the fluctuating hydrodynamic models with four different sets of cutoff parameters.
We assume the values of cutoff parameters $\lambda_\perp$ in the unit of fm and $\lambda_\eta$
are the same for simplicity in the present study.

\begin{table}[htbp]
\caption{Parameters of transport coefficients and initial conditions
 in hydrodynamic models}
\begin{tabular}{p{9em}ccccc} \hline \hline
Model & $\eta/s$ & $\lambda_\perp$ (fm) & $\lambda_\eta$ & $C/\tau_0$ (fm$^{-1})$ & $\alpha$  \\ \hline
Ideal hydro & 0           & N/A & N/A    & 62 & 0.08	\\
Viscous hydro & $1/4\pi$ & N/A & N/A     & 49 & 0.13  \\
Fluc. hydro--$\lambda$2.5 & $1/4\pi$ & 2.5 & 2.5 & 47 & 0.14 \\
Fluc. hydro--$\lambda$2.0 & $1/4\pi$ & 2.0 & 2.0 & 42 & 0.16 \\
Fluc. hydro--$\lambda$1.5 & $1/4\pi$ & 1.5 & 1.5 & 41 & 0.16 \\
Fluc. hydro--$\lambda$1.0 & $1/4\pi$ & 1.0 & 1.0 & 31 & 0.20 \\
\hline \hline
\end{tabular}
\label{table:parameter}
\end{table}
Since the parameters $C$ and $\alpha$ which determine the initial entropy production
are tuned to reproduce the final charged-hadron multiplicity,
their tuned values are sensitive to the characteristics of the entropy production during the hydrodynamic evolution,
specifically, to the presence and magnitudes of the dissipations and fluctuations.
Consequently the overall factor $C/\tau_0$ has a maximum value in the ideal hydrodynamic model
and decreases with increasing magnitude of viscosities and/or fluctuations.

Figure~\ref{fig:multiplicity} shows
the centrality dependence of charged-hadron multiplicity $dN_\mathrm{ch}/d\eta_p$
per participant pair $N_\mathrm{part}/2$ at midrapidity $|\eta_p|<0.5$
in Pb+Pb collisions at $\sqrt{s_{NN}} = 2.76\ \text{TeV}$.
The parameter $C/\tau_0$ and $\alpha$ control the overall magnitude
and the slope of multiplicity per participant pair, respectively.
Within the current framework with two adjustable parameters, $C$ and $\alpha$,
we cannot perfectly reproduce experimental data of multiplicity per participant pair
at the same time in all the ranges of the centrality from central to peripheral collisions.
We tune these two parameters
to reproduce multiplicity at $N_\mathrm{part}\gtrsim150$.
When we compare our results with
experimental ratios of the factorization ratios in Sec.~\ref{sec:Factorization ratio},
we use the events at 0--30\% centrality in which the average number of
participants is above $N_\mathrm{part} \sim 150$.
In fact, the average number of participants
is slightly larger than the one of the experimental data at $N_\mathrm{part}\lesssim200$
and the multiplicity per participant pair is slightly out of error-bars of the experimental data
 at a given centrality below $N_\mathrm{part} \sim 150$.
This requires a more sophisticated initialization method in the hydrodynamic models,
which is beyond the scope of the present paper.

Figures~\ref{fig:multiplicityeta} (a)--(h) show the pseudorapidity
dependence of charged-hadron multiplicity in Pb+Pb collisions at $\sqrt{s_{NN}} = 2.76\ \text{TeV}$
for each centrality.
Experimental data obtained by the ALICE Collaboration~\cite{Abbas:2013bpa,Adam:2015kda}
are reproduced by all hydrodynamic models
in a wide rapidity region in 0--30\% centrality.
The parameters $\sigma_\eta$ and $\Delta\eta$ control the shape of
the distributions.
We choose a single set of these parameters in each hydrodynamic model so as to
reproduce experimental data in central collisions,
which correspond to $N_\mathrm{part}\gtrsim150$ in Fig.~\ref{fig:multiplicity}.
Since we do not reproduce the centrality dependence of charged-hadron multiplicity at midrapidity
below $N_\mathrm{part} \sim 150$ as shown in Fig.~\ref{fig:multiplicity},
pseudorapidity distributions are systematically larger than the experimental data in 30--70\%
regardless of the hydrodynamic models.

Figure~\ref{fig:v2} shows the
transverse momentum ($p_T$) dependence of elliptic flow parameters $v_{2}\{2\}$
of charged hadrons
in 40--50\% central Pb+Pb collisions at $\sqrt{s_{NN}}= 2.76\ \text{TeV}$.
The elliptic flow parameters $v_{2}\{2\}\left(p_T\right)$ from the ideal hydrodynamic model
is systematically larger than experimental data.
Whereas,
the shear viscosity suppresses $v_{2}\{2\}\left(p_T\right)$
in the viscous and the fluctuating hydrodynamic models.
We reproduce the experimental data below $p_{T} = 1.5\ \text{GeV}$
in the viscous  and the fluctuating hydrodynamic models with $\eta/s = 1/4\pi$.
The elliptic flow parameters $v_{2}\{2\}\left(p_T\right)$ from the viscous and the fluctuating hydrodynamic models
are almost the same with each other.
This indicates that the
hydrodynamic fluctuations do not affect $v_{2}\{2\}\left(p_T\right)$ apparently.
Although hydrodynamic fluctuations are supposed to
enhance fluctuations of elliptic flow parameters in general,
its effect  is not significant at this centrality.
Note that we also obtain similar results of $v_{2}\{2\}\left(p_T\right)$
with the other values of $\lambda_\perp$ and $\lambda_\eta$ (not shown).

\subsection{Factorization ratio}
\label{sec:Factorization ratio}
The hydrodynamic fluctuations disturb fluid evolution randomly in space and time
according to
Eq.~\eqref{eq:shear_stress_tensor}.
Therefore the correlations embedded in the initial longitudinal profiles such as
alignment of participant-plane angles along space-time rapidity tend to be broken
due to hydrodynamic fluctuations.
To analyze such effects of the hydrodynamic fluctuations
in the final state observables,
we calculate the factorization ratios~\cite{Gardim:2012im}.

The factorization ratio in the longitudinal direction $r_n(\eta^a_p, \eta^b_p)$ is defined as
\begin{align}
\label{eq:factorization_ratio}
r_n(\eta^a_p, \eta^b_p) &= \frac{V_{n\Delta}(-\eta^a_p, \eta^b_p)}{V_{n\Delta}(\eta^a_p, \eta^b_p)},\\
V_{n\Delta}(\eta^a_p, \eta^b_p)
& = \frac{ \int \cos (n \Delta \phi) \frac{d^2 N}{d\phi^a d\eta_p^a d\phi^b  d\eta_p^b}d\phi^a d\phi^b}{%
    \int \frac{d^2 N}{d\phi^a  d\eta_p^a d\phi^b  d\eta_p^b}d\phi^a d\phi^b}.
\end{align}
Here $V_{n\Delta}(\eta^a_p, \eta^b_p)$
is the Fourier coefficient of two-particle azimuthal correlation functions
at the $n$-th order.
$\Delta \phi = \phi^a - \phi^b$ represents
the difference of azimuthal angles between two charged hadrons.
These two hadrons are taken from the two separated pseudorapidity, $\eta^a_p$ and $\eta^b_p$.

In the event-by-event calculations,
the actual expression of $V_{n\Delta}(\eta^a_p, \eta^b_p)$ is given by
\begin{align}
\label{eq:actual_factorization_ratio}
V_{n\Delta}(\eta^a_p, \eta^b_p)
& = \dlangle{\cos (n\Delta\phi)} \drangle
  = \frac{\Re \langle Q_n^{a*} Q_n^b\rangle}{\langle M^a M^b\rangle},
\end{align}
where $M^{a/b}$ and $Q_n^{a/b}=\sum_i e^{in\phi_i}$ are
the multiplicity and the flow vector in the pseudorapidity bins at $\eta_p^{a/b}$
in a single event.
The single angle brackets $\langle\cdots\rangle$ represent
the average over the events at a given centrality,
while the double angle brackets $\dlangle\cdots\drangle$ represent
the average over the particle pairs in each event
in addition to the event average.

The factorization ratio \eqref{eq:factorization_ratio} can be interpreted as the correlation
between the flows in two pseudorapidity regions, $\eta^a_p$ and  $-\eta^a_p$.
Here the correlation of the event-plane angles (\thatis, $\Psi_n^a$ and $\Psi_n^{-a}$)
and that of the flow magnitudes (\thatis, $v_n^a$ and $v_n^{-a}$)
can be considered separately.
The two-particle correlations $V_{n\Delta}$ \eqref{eq:actual_factorization_ratio} can be naively understood as
\begin{align}
\label{eq:two-part-corr}
  &V_{n\Delta}(\eta^a_p, \eta^b_p) \nonumber \\
  &\quad = \dlangle \cos \left[n(\phi^a-\phi^b)\right]
  \drangle \nonumber \\
  &\quad = \dlangle \cos \{n\left[(\phi^a-\Psi_n^a)-(\phi^b-\Psi_n^b)+(\Psi_n^a-\Psi_n^b)\right]\}
  \drangle \nonumber \\
  &\quad \approx
  \langle v_n^a v_n^b
  \cos [n(\Psi_n^a-\Psi_n^b)] \rangle
\end{align}
by using the event-by-event flow magnitude $v_n^{a/b}= v_n(\eta_p^{a/b})$ and
the event-plane angle $\Psi_n^{a/b} = \Psi_n(\eta^{a/b}_p)$.

If we assume that the event-plane angles are completely aligned along rapidity,
\thatis, $\Psi_n^a=\Psi_n^b$,
and also that the product of flow magnitudes at two different pseudorapidity regions factorized in the event averages,
the two-particle correlations $V_{n\Delta}$ would be simplified:
\begin{align}
  V_{n\Delta}(\eta^a_p, \eta^b_p)
  &\approx \langle v_n^a v_n^b\rangle \approx \langle v_n^a\rangle\langle v_n^b\rangle.
  \label{eq:factorization-1}
\end{align}
In this ideal case, the factorization ratio becomes unity:
\begin{align}
  r_n(\eta^a_p, \eta^b_p)
  \approx \frac{\langle v_n^{-a}\rangle\langle v_n^b\rangle}{\langle v_n^a\rangle\langle v_n^b\rangle} = 1,
\end{align}
since the flow magnitude is symmetric with respect to the pseudorapidity
after averaging over all the events,
\thatis, $\langle v_n^a \rangle= \langle v_n^{-a} \rangle$,
in symmetric systems like Pb+Pb collisions.
However, the event-plane angle depends on pseudorapidity,
therefore the two-particle correlations $V_{n\Delta}$ \eqref{eq:two-part-corr} cannot be factorized
unlike in Eq.~\eqref{eq:factorization-1}.
Nevertheless,
one still assumes the factorization among the flow magnitudes and the event-plane angles.
In this case the two-particle correlation is written as
\begin{align}
  V_{n\Delta}(\eta^a_p, \eta^b_p)
  &\approx \langle v_n^a \rangle \langle v_n^b\rangle
    \langle \cos [n(\Psi_n^a-\Psi_n^b)] \rangle.
\end{align}
Thus the factorization ratio has the following form:
\begin{align}
\label{eq:fact_break_no_fluc}
  r_n(\eta^a_p, \eta^b_p)
  &\approx \frac{
    \langle \cos [n(\Psi_n^{-a}-\Psi_n^b)] \rangle}{
    \langle \cos [n(\Psi_n^a-\Psi_n^b)] \rangle} < 1.
\end{align}
The factorization ratio becomes smaller than unity
since the event-plane angles are expected to be $(\Psi^b - \Psi^{-a}) \ge (\Psi^b - \Psi^a)$ on average
due to the ordering of the rapidity gap $(\eta^b_p - \eta^{-a}_p) \ge (\eta^b_p - \eta^a_p)$.
Therefore, the factorization ratio being smaller than unity implies that
the event-plane angle depends on pseudorapidity.

Besides the event-plane angle decorrelation,
asymmetry of the event-by-event flow magnitude also causes the factorization breakdown.
Even if the event-plane angles are completely aligned,
the two-particle correlations have an additional term
coming from the correlation of the flow magnitude fluctuations:
\begin{align}
  V_{n\Delta}(\eta^a_p, \eta^b_p)
  &\approx \langle v_n^a v_n^b\rangle
  = \langle v_n^a\rangle\langle v_n^b\rangle + \langle \Delta v_n^a \Delta v_n^b\rangle,
\end{align}
where $\Delta v_n$ is the event-by-event flow magnitude fluctuations:
\begin{align}
v_n = \langle v_n\rangle + \Delta v_n.
\end{align}
In this case, the factorization ratio becomes
\begin{align}
\label{eq:fact_break_full}
  r_n(\eta^a_p, \eta^b_p) &\approx
  \frac{\langle v_n^{-a}\rangle \langle v_n^b\rangle + \langle \Delta v_n^{-a} \Delta v_n^b\rangle}
  {\langle v_n^a\rangle \langle v_n^b\rangle + \langle \Delta v_n^a \Delta v_n^b\rangle}.
\end{align}
If the flow magnitude fluctuates asymmetrically in different pseudorapidity regions,
$\langle \Delta v_n^{-a} \Delta v_n^b\rangle \le \langle \Delta v_n^a \Delta v_n^b\rangle$
is expected due to the ordering of the rapidity gap. Thus the denominator becomes larger than the numerator in Eq.~\eqref{eq:fact_break_full}.
Therefore the factorization ratio \eqref{eq:fact_break_full} becomes smaller than unity,
$r_n(\eta^a_p, \eta^b_p)<1$,
due to the reason similar to the case of the event-plane angle decorrelation
in Eq.~\eqref{eq:fact_break_no_fluc}.

To summarize, the factorization ratio being smaller than unity indicates that
the event-by-event rapidity decorrelation of
the event-plane angle and/or the flow magnitude,
\thatis, the event-plane angle $\Psi_n$ and/or the flow magnitude $v_n$
depend on pseudorapidity in an event.
Further discussions will be given in Sec.~\ref{sec:Legendre coefficients}
by introducing the Legendre coefficients
to discriminate
between the event-plane decorrelation and flow magnitude decorrelation.

In the following,
$V_{n\Delta}(\eta^a_p, \eta^b_p)$
is calculated with
the rapidity region $0<\eta^a_p<2.5$ and
the transverse momentum region $0.3<p_T^a<3.0$ for a particle $a$, and
the reference rapidity region $3.0<\eta^b_p<4.0$ for a  particle $b$.
These rapidity and transverse momentum regions
follow the experimental setup of the CMS Collaboration~\cite{Khachatryan:2015oea}.

\begin{figure}[htbp]
\centering
\includegraphics[width=0.5\textwidth,  bb=60 0 320 220]{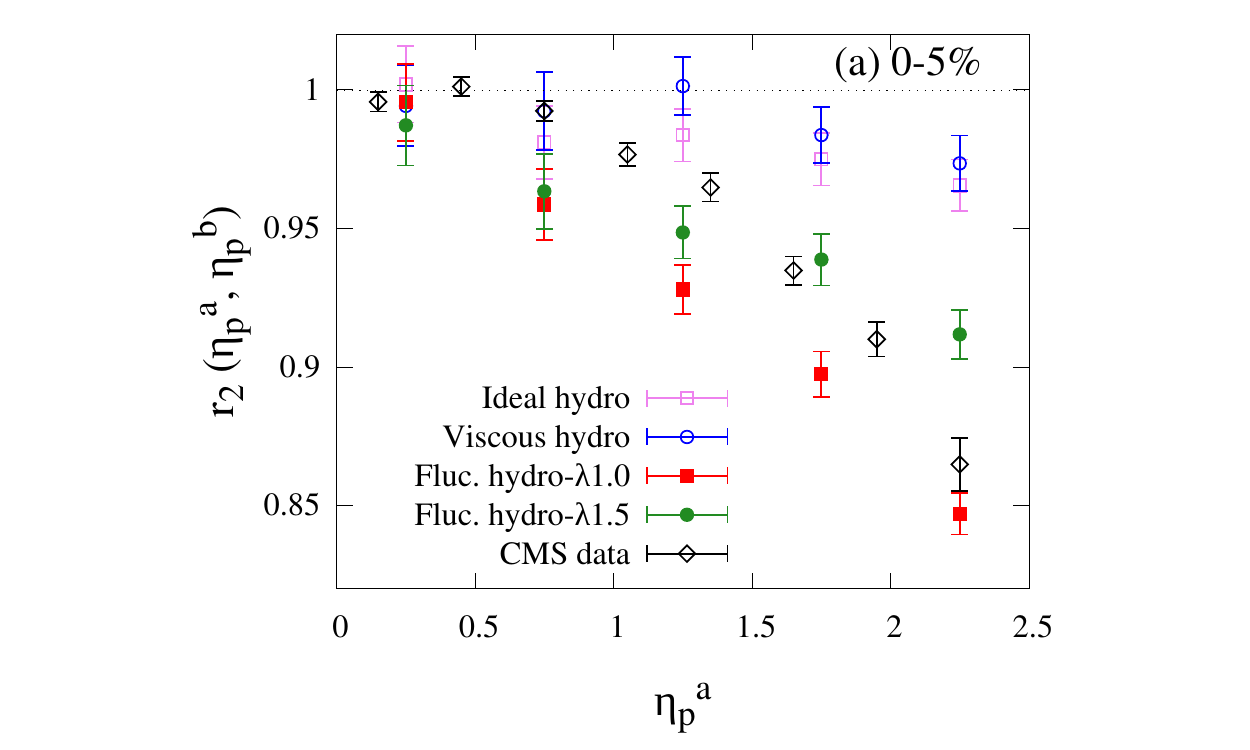}
\includegraphics[width=0.5\textwidth,  bb=60 0 320 220]{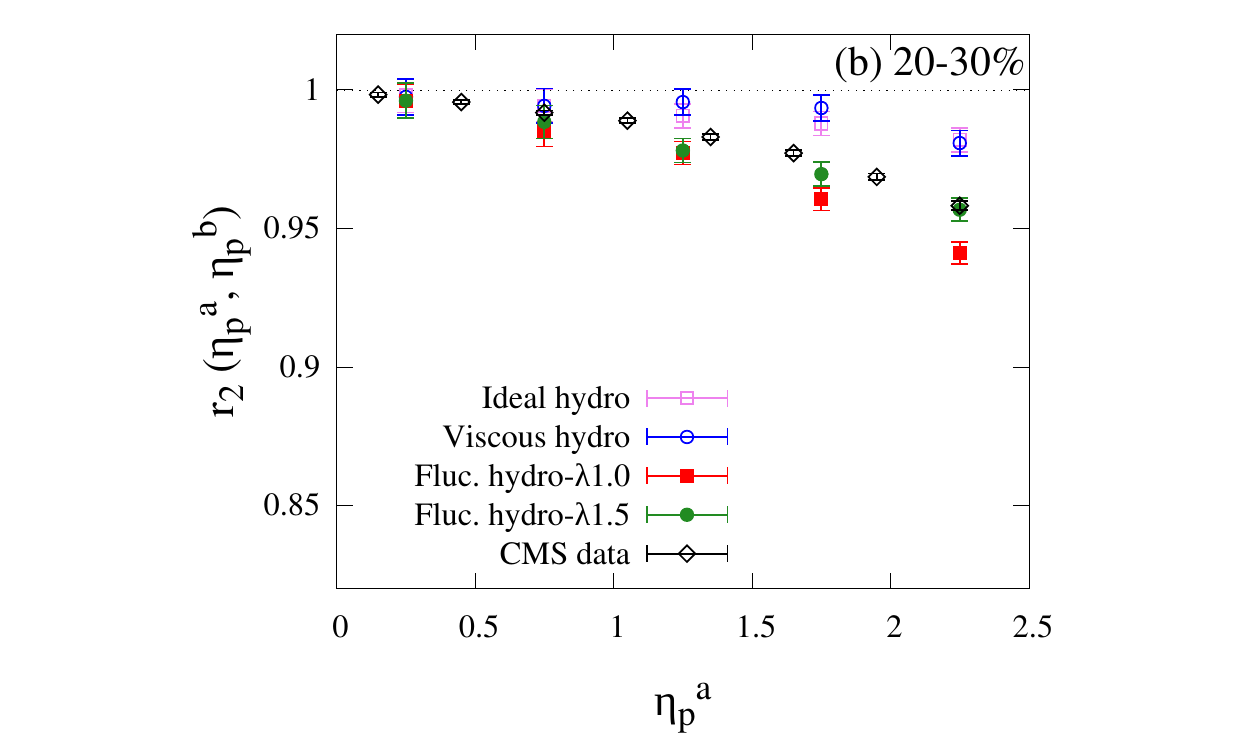}
\caption{(Color Online)
Factorization ratio in the longitudinal direction, $r_2(\eta^a_p,\eta^b_p)$,
in Pb+Pb collisions
at $\sqrt{s_{NN}} = 2.76$ TeV for (a) 0--5\% and (b) 20--30\%
centralities.
The rapidity region for reference is  $3.0<\eta_p^b<4.0$.
The results from ideal hydrodynamics (open square), viscous hydrodynamics (open circle),
fluctuating hydrodynamics--$\lambda1.0$ (filled square)
and fluctuating hydrodynamics--$\lambda1.5$ (filled circle)
are compared with experimental data (open diamond) obtained by the CMS Collaboration~\cite{Khachatryan:2015oea}.}
\label{fig:r2}
\end{figure}

Figure~\ref{fig:r2} shows the factorization ratio in the longitudinal direction,
$r_2(\eta^a_p, \eta^b_p)$,
in Pb+Pb collisions at $\sqrt{s_{NN}} = 2.76$ TeV
for 0--5\% and 20--30\% centralities.
The experimental data of $r_2(\eta^a_p, \eta^b_p)$
from the CMS Collaboration \cite{Khachatryan:2015oea}
is close to unity at small $\eta^a_p$
and decreases with increasing $\eta^a_p$.
On the other hand, the factorization ratios $r_2(\eta^a_p, \eta^b_p)$
from the ideal and viscous hydrodynamic models modestly decrease with increasing $\eta^a_p$
in comparison with the experimental data, which
means that expansion of the fluids tends to keep the long-range correlation
in the rapidity direction.
As discussed in Sec.~\ref{sec:initial_condition},
the initial entropy distributions \eqref{eq:N}
are smooth functions of space-time rapidity and
their transverse profile fluctuates due to the random positions of nucleons in colliding nuclei.
In such a case, its participant-plane angles are almost the same for different space-time rapidities.
Therefore, ideal and viscous hydrodynamic evolution directly translates longitudinal correlations
in the initial profiles
into correlations of the event-plane angle along pseudorapidity in final momentum anisotropy.
It should be noted here that the small decorrelation in ideal and viscous hydrodynamic models
is understood from the weak twist structure in the modified BGK model which is mentioned
in Sec.~\ref{sec:initial_condition}.
In comparison with the results from the ideal and the viscous hydrodynamic models,
the factorization ratios from the fluctuating hydrodynamic models
are significantly smaller than unity at a large rapidity separation
and are comparable with the experimental data.
This indicates that hydrodynamic fluctuations
break the factorization of the two-particle correlation function
and cause rapidity decorrelation of the magnitude of anisotropic flow and/or the event-plane angle.
Results from the fluctuating hydrodynamic models depend on the cutoff parameters
in Eq.~\eqref{eq:Gaussian_with_cutoff}.
Larger cutoff parameters correspond to smaller magnitudes of hydrodynamic fluctuations,
and fluctuating hydrodynamics is reduced to the viscous hydrodynamics in the large limit of these parameters.
The results shown in Fig.~\ref{fig:r2} are consistent with this perspective.

\begin{figure}[htbp]
\centering
\includegraphics[width=0.5\textwidth,  bb=60 0 320 220]{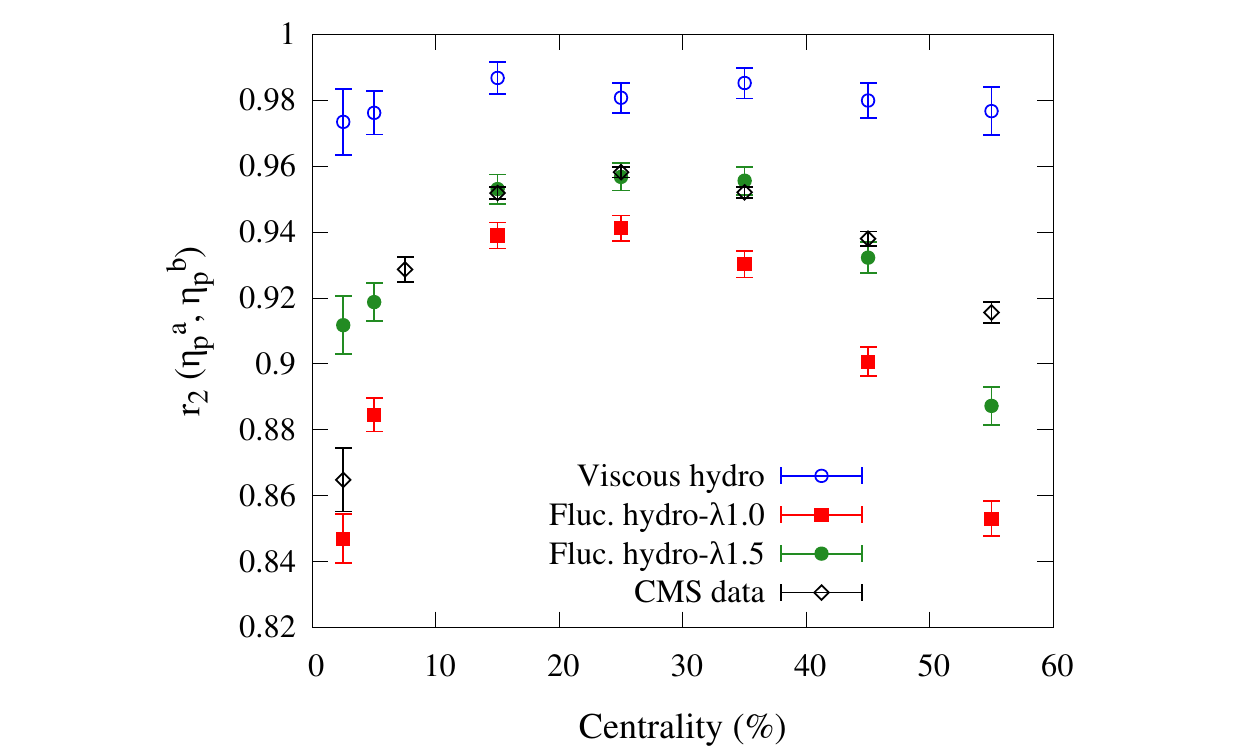}
\caption{(Color Online)
Centrality dependence of factorization ratio $r_2(\eta^a_p, \eta^b_p)$
in Pb+Pb collisions at $\sqrt{s_{NN}} = 2.76$ TeV.
The rapidity regions of two-particle correlation functions are
taken as $2.0<\eta^a_p<2.5$ and $3.0<\eta^b_p<4.0$.
The symbols are the same as in Fig.~\ref{fig:r2}.
The experimental data are taken from Ref.~\cite{Khachatryan:2015oea}.
}
\label{fig:r2comp}
\end{figure}

Figure~\ref{fig:r2comp} shows the centrality dependence of $r_2(\eta^a_p, \eta^b_p)$
in Pb+Pb collisions at $\sqrt{s_{NN}} = 2.76$ TeV.
The experimental data of $r_2(\eta^a_p, \eta^b_p)$ are smaller in the central collisions (0--10\% centrality)
and the peripheral collisions (50--60\% centrality) than in the semi-central collisions
(10--50\% centrality).
This is understood from the correlations embedded in the initial profiles.
In the semi-central collisions, the initial geometry becomes elliptical
to generate strong correlations
of the participant-plane angle in the longitudinal direction.
This initial correlation is reflected in the final momentum anisotropy, and
the factorization ratio $r_2(\eta^a_p, \eta^b_p)$ tends to be close to unity in semi-central collisions.
The factorization ratio $r_2(\eta^a_p, \eta^b_p)$ from the viscous hydrodynamic model
is close to unity and this model fails to reproduce the experimental data
for the whole centrality region.
In contrast,
$r_2(\eta^a_p, \eta^b_p)$ obtained from the fluctuating hydrodynamic models
show qualitatively the same behavior as experimental data.
In the central collisions (0--10\% centrality), the fluctuating hydrodynamic model--$\lambda1.0$
reproduces the experimental data reasonably well.
Whereas, the fluctuating hydrodynamic model--$\lambda1.5$ better
reproduces the experimental data in the semi-central collisions (centrality classes between 10--50\%).
Therefore, the fluctuating hydrodynamic model
could reproduce the experimental data of
the centrality dependence of $r_2(\eta^a_p, \eta^b_p)$
by using a different set of the Gaussian width (\thatis, $\lambda_\perp$ and $\lambda_\eta$)
for a different centrality.

\begin{figure}[htbp]
\centering
\includegraphics[width=0.5\textwidth,  bb=60 0 320 220]{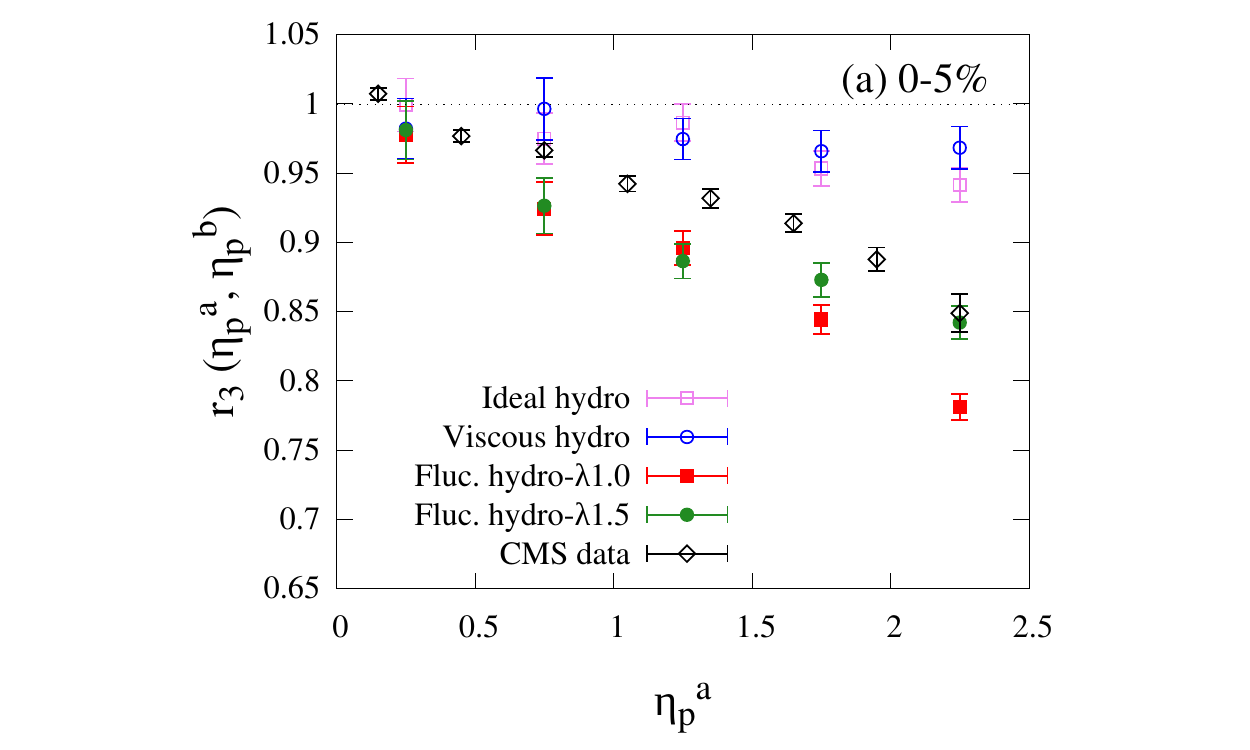}
\includegraphics[width=0.5\textwidth,  bb=60 0 320 220]{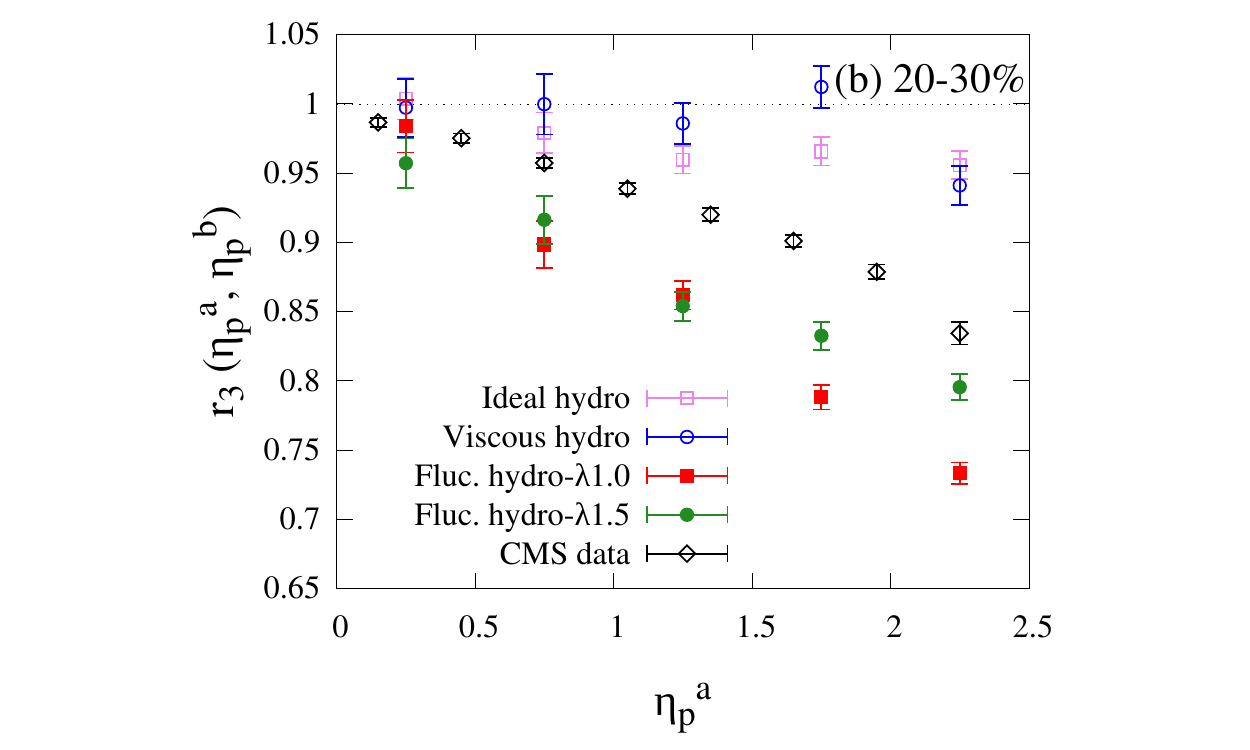}
\caption{(Color Online)
Factorization ratio $r_3(\eta^a_p,\eta^b_p)$ in Pb+Pb collisions
at $\sqrt{s_{NN}} = 2.76\ \text{TeV}$ for  (a) 0--5\% and (b) 20--30\%
centralities.
The symbols are the same as in Fig.~\ref{fig:r2}.
The experimental data are taken from Ref.~\cite{Khachatryan:2015oea}.
}
\label{fig:r3}
\end{figure}

The second-order anisotropic flow (elliptic flow)
is driven mainly by the initial geometry.
In contrast, the third- (or, in general, odd-) order anisotropic flows
are purely caused by fluctuations.
The pattern of fluctuations in the initial transverse profile originating
from the configuration of nucleons in the colliding nuclei
is almost the same for different rapidities.
The event-plane angles at the third order would also be correlated along pseudorapidity
as in the case for elliptic flow.
To suppress the effect of initial collision geometry on the factorization ratios
and to see the effects of hydrodynamic fluctuations more directly,
we also analyze $r_3(\eta^a_p, \eta^b_p)$.
The third-order anisotropic flow (triangular flow)
 at a given pseudorapidity is generated from
fluctuations of initial transverse profile at almost the same space-time rapidity.
Figure~\ref{fig:r3} shows the factorization ratio $r_3(\eta^a_p, \eta^b_p)$
in Pb+Pb collisions at $\sqrt{s_{NN}} = 2.76$ TeV for 0--5\% and 20--30\% centralities.
The experimental data of $r_3(\eta^a_p, \eta^b_p)$
from the CMS Collaboration \cite{Khachatryan:2015oea}
decreases with increasing $\eta_p^a$.
On the other hand,
$r_3(\eta^a_p, \eta^b_p)$ from the viscous hydrodynamic model are close to unity
and almost the same as those from the ideal hydrodynamic model.
This indicates the viscosity itself does not further affect rapidity decorrelation of the third-order
anisotropic flow.
Whereas, hydrodynamic fluctuations reduce $r_3(\eta^a_p, \eta^b_p)$ considerably
and it drops linearly with $\eta_p^a$.
Although the experimental data of factorization ratio $r_2(\eta^a_p, \eta^b_p)$ are roughly
reproduced with $\lambda_\perp  = 1.0$--$1.5\ \text{fm}$ in the fluctuating hydrodynamic model,
$r_3(\eta^a_p, \eta^b_p)$ with the same setups are systematically smaller than the experimental data.
Note that this is the opposite trend seen by the twist structure of the initial profiles in Ref.~\cite{Bozek:2017qir}
in which $r_2$ decorrelates too much while $r_3$ reproduces experimental data.

\begin{figure}[htbp]
\centering
\includegraphics[width=0.5\textwidth,  bb=60 0 320 220]{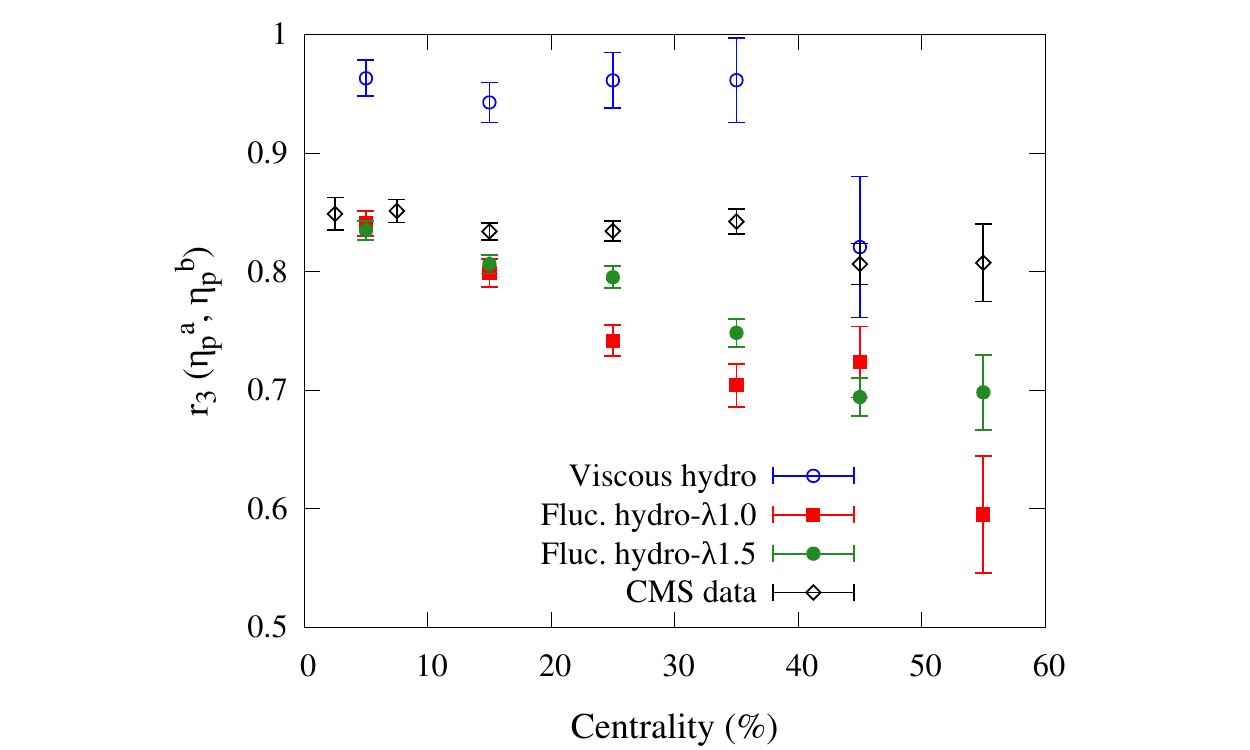}
\caption{(Color Online)
Centrality dependence of factorization ratio $r_3(\eta^a_p, \eta^b_p)$
in Pb+Pb collisions at $\sqrt{s_{NN}} = 2.76$ TeV.
The rapidity regions of two-particle correlation functions are
taken as $2.0<\eta^a_p<2.5$ and $3.0<\eta^b_p<4.0$.
The meaningful result for 50--60\% centrality from viscous hydrodynamic model
was not obtained due the insufficient statistics.
The symbols are the same as in Fig.~\ref{fig:r2}.
The experimental data are taken from Ref.~\cite{Khachatryan:2015oea}.
}
\label{fig:r3comp}
\end{figure}

Figure~\ref{fig:r3comp} shows the centrality dependence of $r_3(\eta^a_p, \eta^b_p)$
in Pb+Pb collisions at $\sqrt{s_{NN}} = 2.76$ TeV.
The experimental data has almost no dependence on the centrality.
Since Pb+Pb collision system is symmetric in rapidity,
the longitudinal decorrelation of triangular flows purely comes from fluctuations.
The factorization ratio $r_3(\eta^a_p, \eta^b_p)$ from the viscous hydrodynamic model
is close to unity and this model fails to reproduce the experimental data in central collisions.
In contrast,
$r_3(\eta^a_p, \eta^b_p)$ obtained from the fluctuating hydrodynamic models
reproduces the experimental data reasonably well in the central collisions (0--10\% centrality).
Whereas, $r_3(\eta^a_p, \eta^b_p)$ obtained from the fluctuating hydrodynamic models
are smaller than the experimental data in the peripheral collisions (10--60\% centralities).

\subsection{Legendre coefficients}
\label{sec:Legendre coefficients}
It is known that
the effects of both the flow magnitude asymmetry and the event-plane twist reduce
the factorization ratios~\cite{Jia:2014vja}.
The original definition of the factorization ratio by CMS~\cite{Khachatryan:2015oea}
cannot discriminate between
the effects of the flow magnitude asymmetry and the event-plane twist.
While, improved definitions of factorization ratios are proposed by ATLAS~\cite{Aaboud:2017tql} to discriminate
between them by assuming the linear dependence of flow magnitude and event-plane angles.
In addition to these two rapidity-decorrelation mechanisms,
the analysis in the previous subsection reveals that hydrodynamic fluctuations also reduce
the factorization ratios.
In this subsection, we calculate the Legendre coefficients of the flow magnitude
and the event-plane angle to separately
estimate the effects of the flow magnitude asymmetry
and the event-plane twist.
The Legendre coefficients were previously used to quantify
the longitudinal multiplicity fluctuations~\cite{Monnai:2015sca}.
We show that they can also be used to quantify
the event-by-event longitudinal
structure of the anisotropic flow parameter $v_{n}(\eta_p)$
and the event-plane angle $\Psi_n(\eta_p)$.

The pseudorapidity dependence of anisotropic flow $v_n$ and its event-plane
angle $\Psi_n$ for each hydrodynamic event are expanded by using the Legendre polynomial $P_k$ as
\begin{align}
    v_n(\eta_p) &= \sum_{k=0}^\infty a_n^k P_k\left(\frac{\eta_p}{\eta^\mathrm{max}_p}\right), \\
    \Psi_n(\eta_p) &= \sum_{k=0}^\infty b_n^k P_k\left(\frac{\eta_p}{\eta^\mathrm{max}_p}\right),
\end{align}
where $a_n^k$ and $b_n^k$ are the Legendre coefficients
which measure the magnitude of the Legendre mode in the longitudinal direction
for the flow parameters and event-plane angles, respectively.
In particular, the magnitudes of $a_n^1$ and $b_n^1$ correspond to
the anisotropic flow asymmetry and the event-plane twist~\cite{Jia:2014vja}, respectively.

These Legendre coefficients fluctuate from event to event
due to the event-by-event twisted initial conditions
and the hydrodynamic fluctuations.
To obtain the magnitudes of these coefficients,
we define the root mean square of the coefficients as
\begin{align}
  A_n^k &= \sqrt{\langle(a_n^k)^2\rangle_{\mathrm{ev}}},&
  B_n^k &= \sqrt{\langle(b_n^k)^2\rangle_{\mathrm{ev}}}.
\end{align}
Here $\langle\cdots\rangle_{\mathrm{ev}}$ represents the average over hydrodynamic events at a given centrality.

We calculate $v_n(\eta_p)$ and $\Psi_n(\eta_p)$ of each hydrodynamic event
combining the final-state hadrons
from all the oversampled cascade events.
Thus we suppress the non-flow effects and the finite particle number effects in this analysis.
We consider a rapidity range $|\eta_p| < \eta^{\mathrm{max}}_p=2.5$
which we used in the analysis of the factorization ratios.
In the following, we consider up to $k=2$
and the first three Legendre polynomials are given as
\begin{align}
  P_0(x) &= 1, &
  P_1(x) &= x, &
  P_2(x) &= \frac{1}{2}(3x^2-1).
\end{align}

\begin{figure}[htbp]
\centering
\includegraphics[width=0.45\textwidth, bb=50 0 320 220]{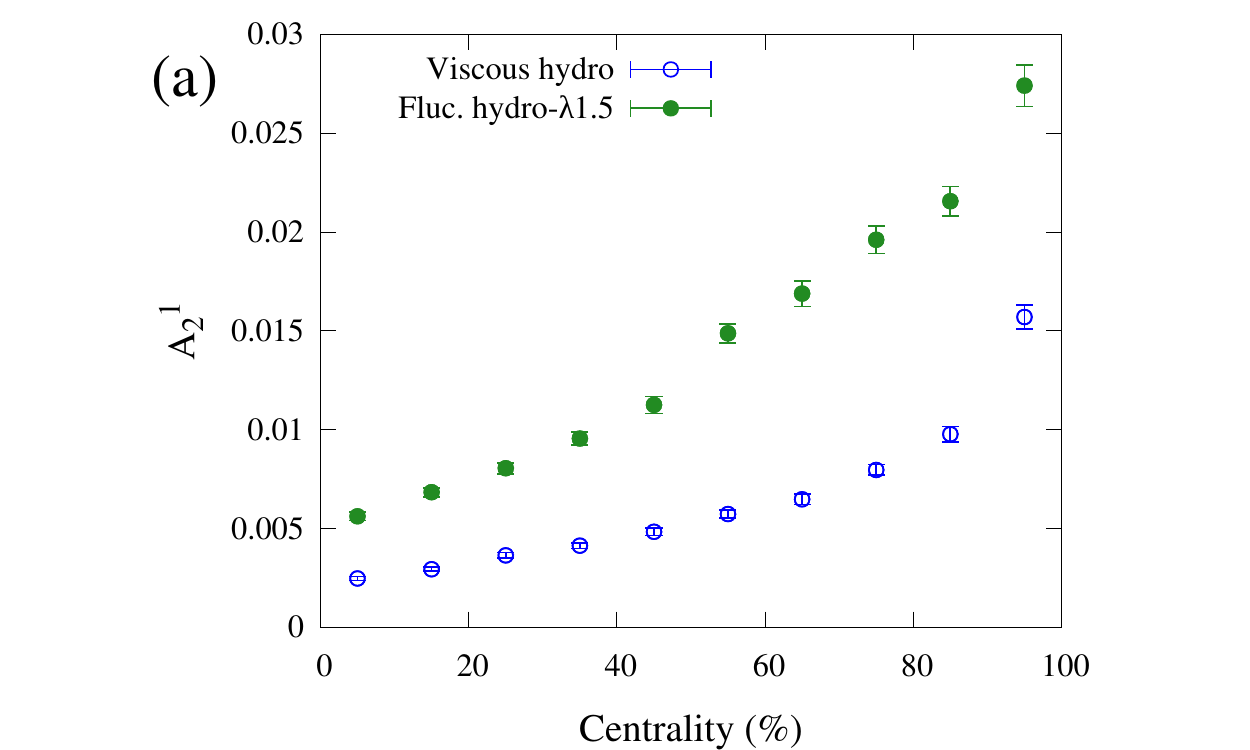}\\
\includegraphics[width=0.45\textwidth, bb=50 0 320 220]{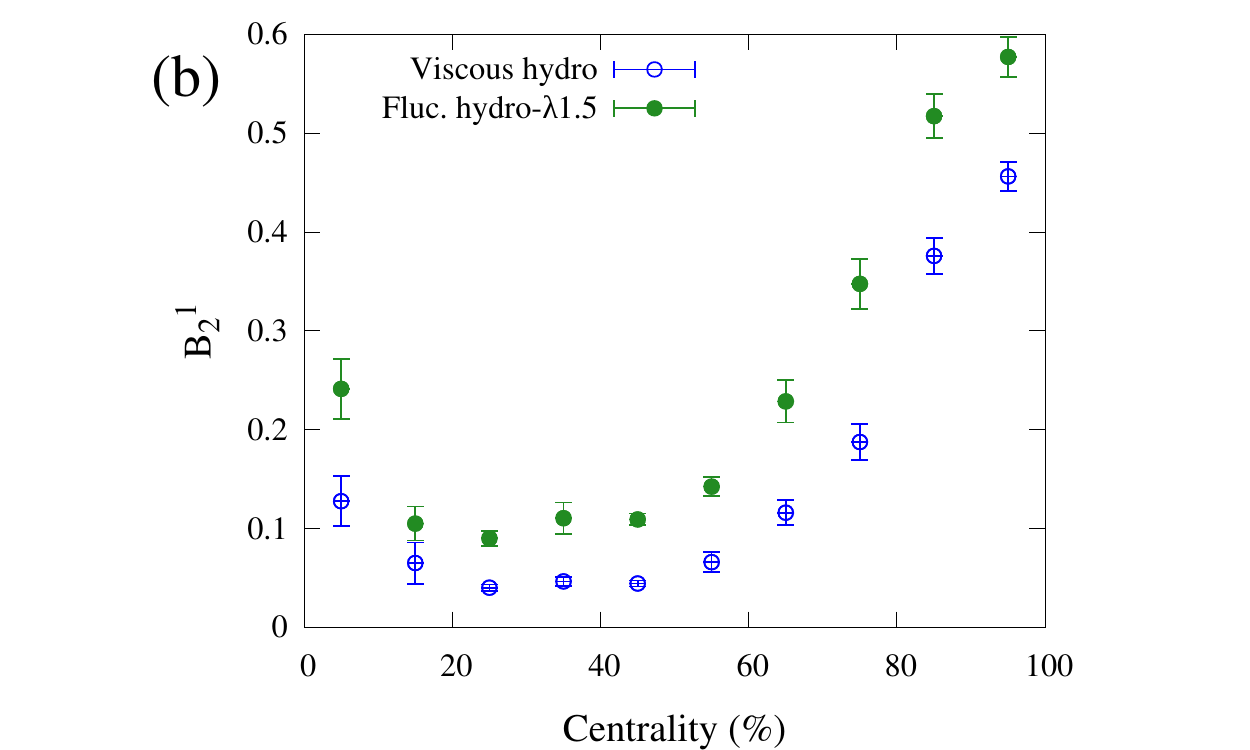}
\caption{(Color Online)
  Centrality dependence of the first-order Legendre coefficients (a) $A^1_2$
  and (b) $B^1_2$ for elliptic flow ($n=2$).
  The results from the viscous hydrodynamic model (open circle)
  and the fluctuating hydrodynamic model (filled circle) are shown for comparison.
}
\label{fig:A1B1}
\end{figure}

Figure~\ref{fig:A1B1} shows the first-order Legendre coefficients $A^1_2$
and $B^1_2$ for elliptic flow ($n = 2$) as functions of centrality
from the viscous and the fluctuating hydrodynamic models.
Both $A^1_2$ and $B^1_2$ from the fluctuating hydrodynamic model are
systematically larger than the ones from
the viscous hydrodynamic model.
This implies that hydrodynamic fluctuations enhance a linear $\eta_p$-dependence
of both the second-order anisotropic flow parameter $v_2$ and its event-plane angle $\Psi_2$ from event to event.
The coefficient $B_2^1$ takes a minimum around 20--50\% for both viscous and fluctuating hydrodynamics.
This is because the event-plane angle is stabilized
by the strong elliptical geometry of initial transverse profiles  in the semi-central collisions.
Therefore, the event-plane angle receives relatively smaller influences
from the twist of the initial conditions and the hydrodynamic fluctuations.
Meanwhile, the coefficient $A_2^1$ monotonically increases as a function of centrality percentile.
This indicates that the development of the flow asymmetry in the longitudinal direction
is independent of the transverse anisotropy driven by collision geometry.
This  increasing  behavior of $A_2^1$
can be understood in the following way:
The longitudinal asymmetries are originally introduced locally at each transverse position
by the rapidity dependence of Eq.~\eqref{eq:N} and hydrodynamic fluctuations
and then reflected in the observed flow by the matter evolution.
When the transverse area of the created matter is large,
such local asymmetry fluctuations are averaged out within the transverse plane
so that the effect on the flow asymmetry fluctuations becomes relatively smaller.
While, in the peripheral collisions,
the local asymmetries can be
directly reflected in the final flow asymmetry
without the averaging effect.
Thus, the magnitude of fluctuations becomes relatively larger in peripheral collisions.

Although the effects of the linear dependence of $v_n(\eta_p)$ and $\Psi_n(\eta_p)$
on the rapidity decorrelation
have been discussed~\cite{Jia:2014vja,Aaboud:2017tql},
discussion on the effects of the higher-order dependence is absent.
From the non-linear behavior of the factorization ratio
shown in the previous subsection,
the higher-order dependence cannot be neglected in understanding the detailed mechanism of
rapidity decorrelation.
Therefore, we also calculate the second-order Legendre coefficients
to quantify non-linear behaviors of the second-order anisotropic flow
as a function of pseudorapidity.

\begin{figure}[htbp]
\centering
\includegraphics[width=0.45\textwidth, bb=50 0 320 220]{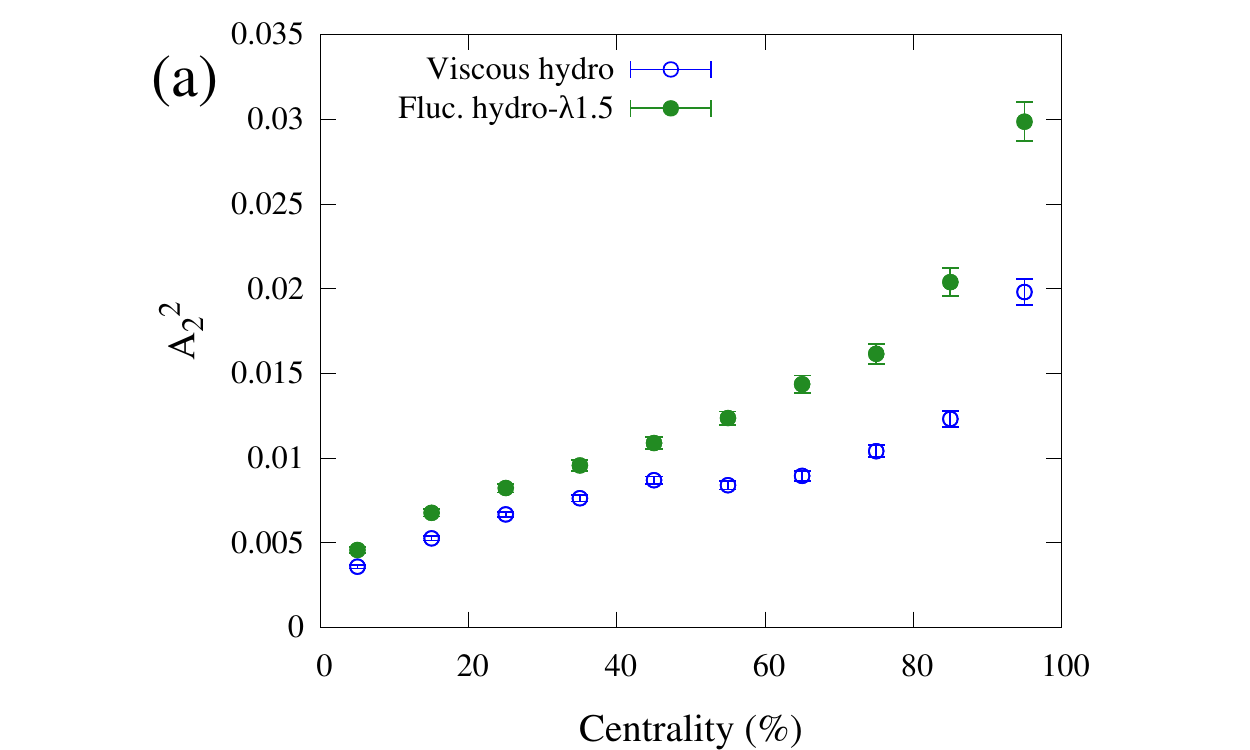}\\
\includegraphics[width=0.45\textwidth, bb=50 0 320 220]{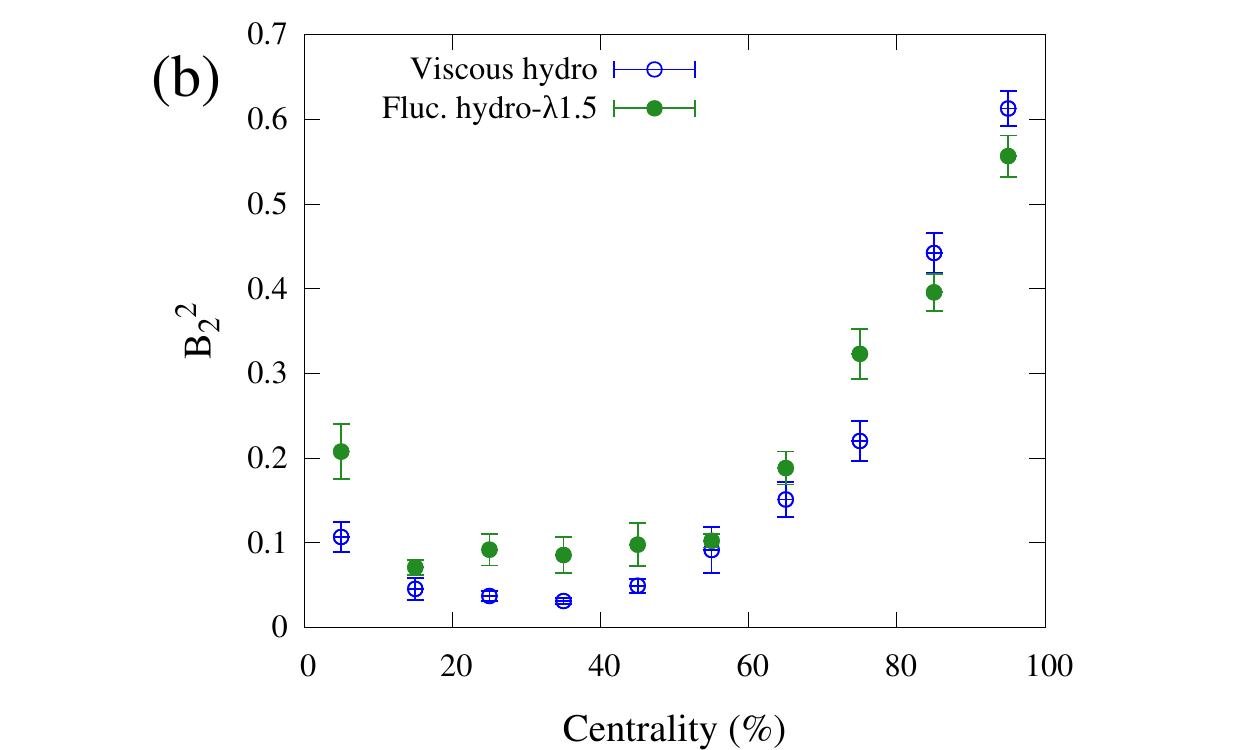}
\caption{(Color Online)
Centrality dependence of the second-order Legendre coefficients (a) $A^2_2$
and (b) $B^2_2$ for elliptic flow ($n=2$).
The symbols are the same as in Fig.~\ref{fig:A1B1}.
}
\label{fig:A2B2}
\end{figure}

Figure~\ref{fig:A2B2} shows the second-order Legendre coefficients $A^2_2$
and $B^2_2$ for elliptic flow ($n = 2$) as a function of centrality from the viscous and
the fluctuating hydrodynamic models.
$A^2_2$ from the fluctuating hydrodynamic model is
systematically larger than the one from
the viscous hydrodynamic model.
$A^2_2$ from viscous hydrodynamics has a slight hump in centrality 30--50\%,
which can be understood from the large magnitude of $v_2$ in this centrality
by considering the fact that the mode $A^2_2$ is  due not only to the flow fluctuations
but also to the average flow magnitude.
While, $A^2_2$ monotonically increases with centrality percentile
in the fluctuating hydrodynamic model,
which implies the dominance of the flow fluctuations over the average flow magnitude.
$B^2_2$ from viscous hydrodynamics
and fluctuating hydrodynamics are both non-zero and even have a similar magnitude to the first-order coefficient $B_2^1$.
In particular, $B^2_2$ are sizable in central (0--10\%) and peripheral (60--100\%) collisions.
Since $B^2_2$ is the Legendre coefficient of the second-order Legendre polynomials,
the event-plane angle is not only linear but quadratic as a function of pseudorapidity.
This means that the event-by-event event-plane angle contains a rapidity-even component and
that the twist direction of the event-plane rotation is not necessarily to be in one direction
unlike in the twist case.
This second-order mode is  larger
in central (0--10\%) and peripheral (60--100\%) than that in semi-central (10--60\%) collisions.
Similar to the first-order Legendre coefficients,
this decrease of $B^2_2$ in semi-central collisions
can be understood from the stabilized event plane
due to the collision geometry.

\begin{figure*}[htbp]
\centering
\begin{tabular}{c@{\hspace{1em}}c}
\includegraphics[width=0.45\textwidth, bb=50 0 320 220]{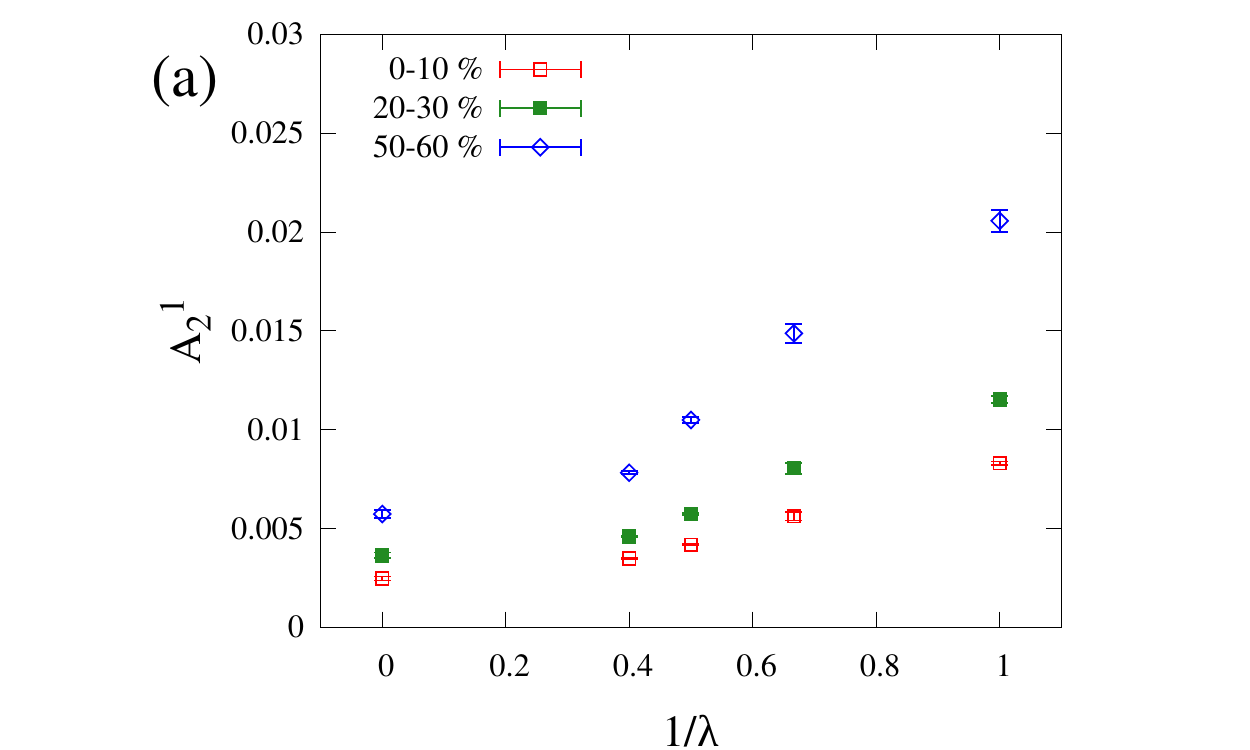} &
\includegraphics[width=0.45\textwidth, bb=50 0 320 220]{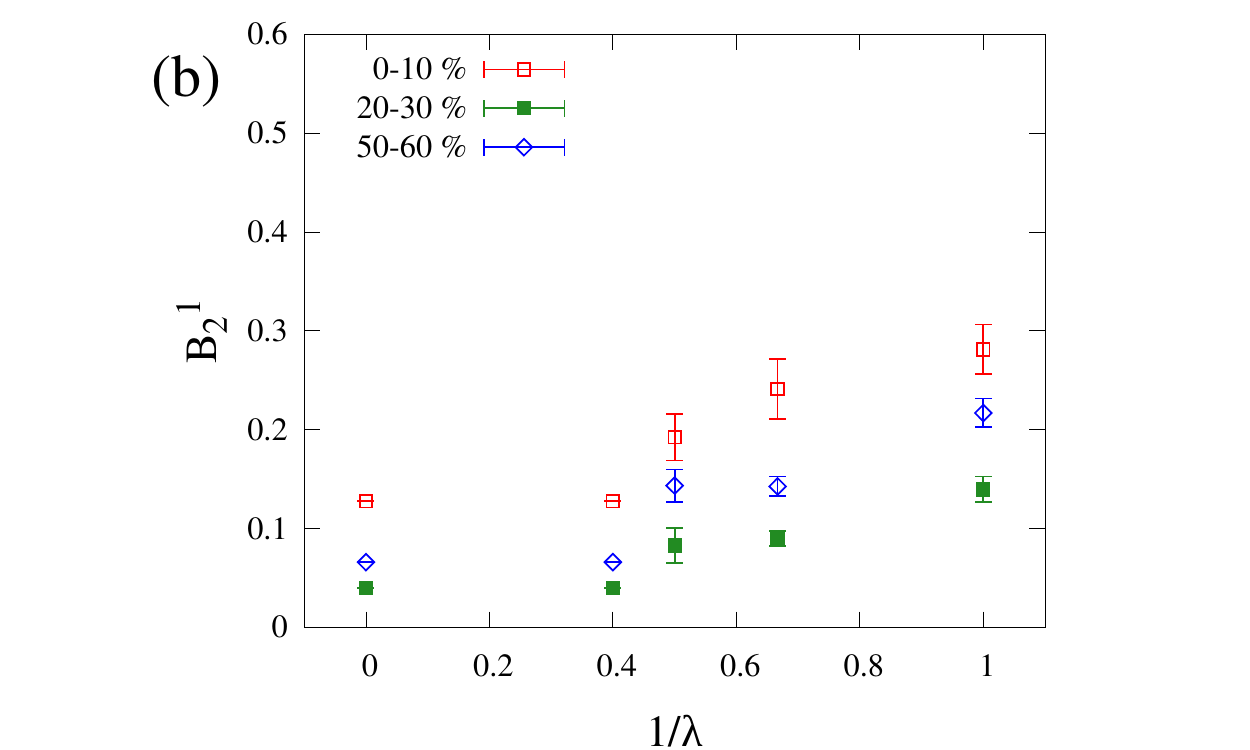} \\
\includegraphics[width=0.45\textwidth, bb=50 0 320 220]{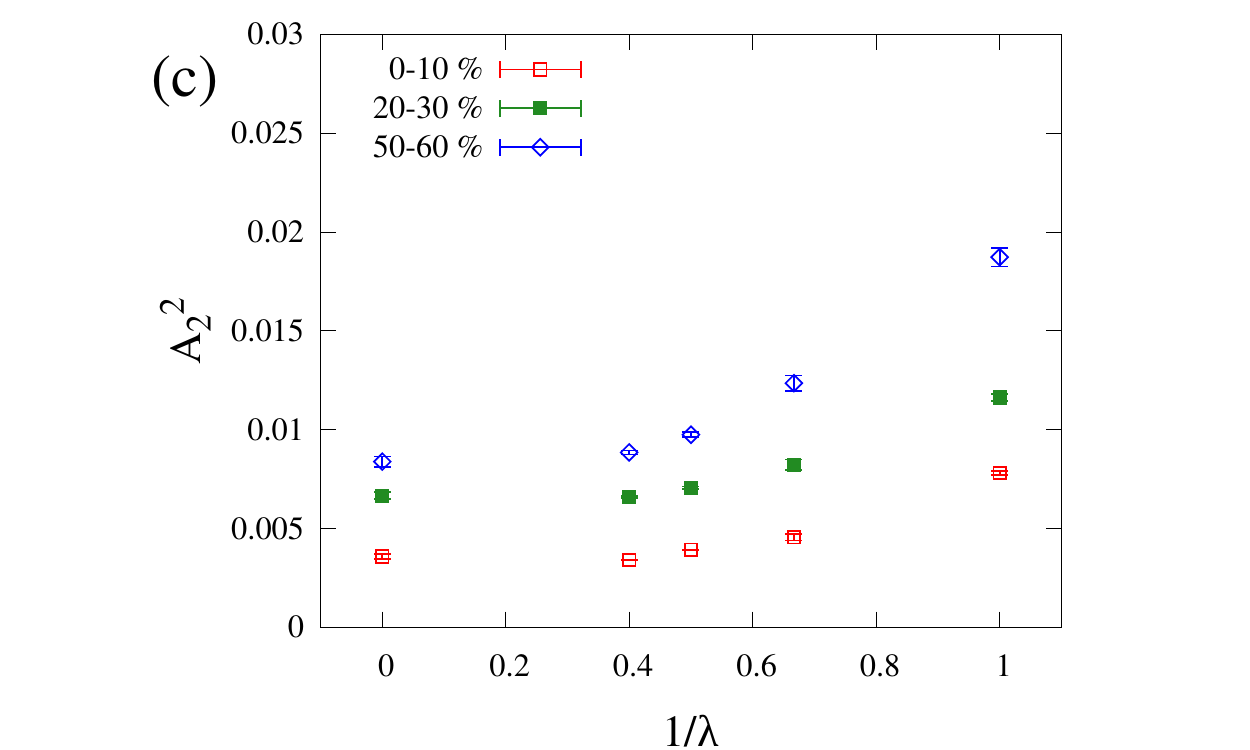} &
\includegraphics[width=0.45\textwidth, bb=50 0 320 220]{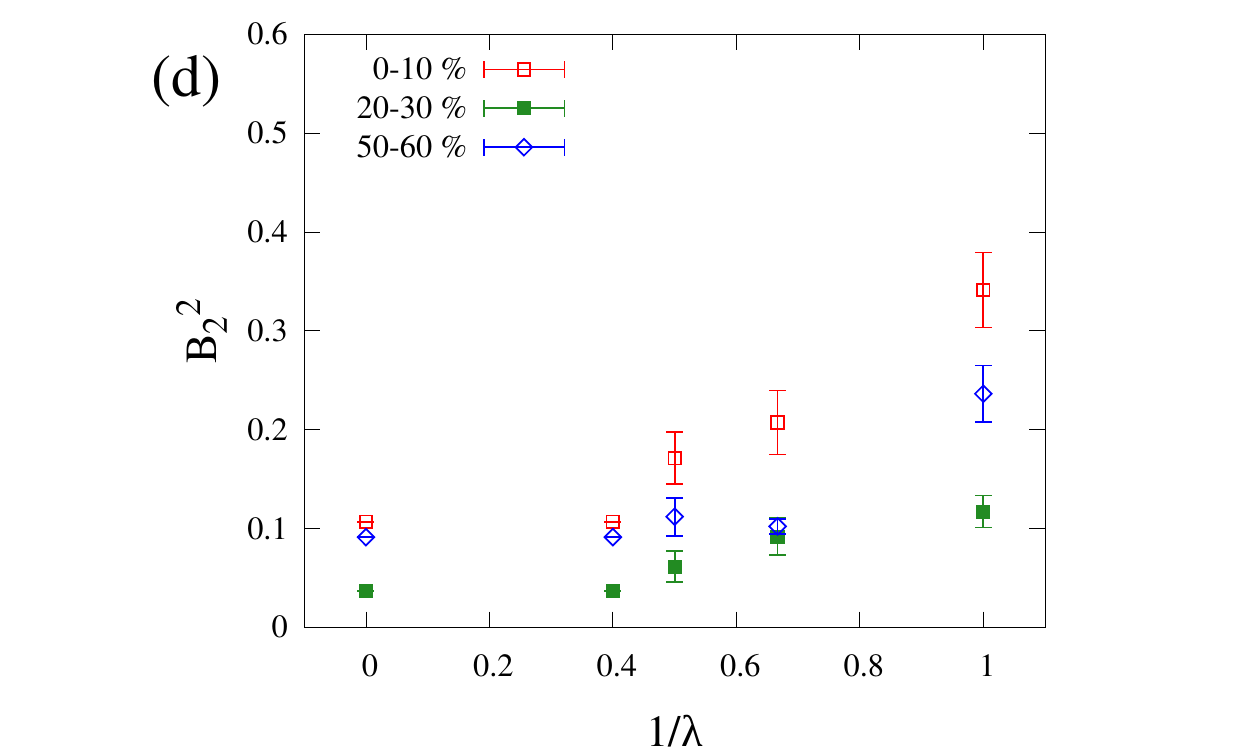}
\end{tabular}
\caption{(Color Online)
Cutoff parameter dependence of the Legendre coefficients.
The first and the second-order Legendre coefficients for the magnitude and the event-plane angle
of elliptic flow parameters, (a) $A^1_2$, (b) $A^2_2$, (c) $B^1_2$, and (d) $B^2_2$,
from viscous hydrodynamics and fluctuating
hydrodynamics--$\lambda1.0$, $\lambda1.5$, $\lambda2.0$, and $\lambda2.5$
are shown.
The results from viscous hydrodynamics are plotted at $1/\lambda=0$
(or $\lambda=\infty$).
The results are shown for centrality 0--10\% (open square), 20--30\% (filled square),
and 50--60\% (open diamond).}
\label{fig:legendrelambda}
\end{figure*}
We analyze the cutoff parameter dependence of the Legendre coefficients
to understand how much the hydrodynamic fluctuations
affect the pseudorapidity dependences of
the magnitude $v_2$ and the event-plane angle $\Psi_2$ of
the second-order anisotropic flow parameters.
Figure~\ref{fig:legendrelambda}
shows the $1/\lambda$ dependence of the first- and second-order
Legendre coefficients from the viscous
and the fluctuating hydrodynamic models.
Here the viscous hydrodynamic model can be regarded as a fluctuating hydrodynamic model
with $\lambda=\infty$ (or $1/\lambda=0$).
The magnitude of hydrodynamic fluctuations becomes larger with the smaller $\lambda$
so that the Legendre coefficients increase
with decreasing $\lambda$ in Fig.~\ref{fig:legendrelambda}.
The coefficients for the magnitude, $A_2^1$ and $A_2^2$,
are minimum in central collisions (0--10\%)
and increase with centrality percentile.
On the other hand, the coefficient for the event-plane angle, $B_2^1$ and $B_2^2$, are minimum in centrality 20--30\%
and larger in central collisions (0--10\%)
and peripheral collisions (50--60\%).
This difference is due to the same reason discussed for Fig.~\ref{fig:A1B1}.
In particular, a monotonically increasing behavior of $A_2^1$ can also
be seen in the second-order Legendre coefficient $A_2^2$.
One sees that
the values of the Legendre coefficients at $1/\lambda = 0.4 = 1.0 / 2.5$
are already almost the same as the ones from the viscous hydrodynamics ($1/\lambda = 0$)
except for $A_2^1$.
Therefore we conclude that hydrodynamic fluctuations
become significant for the physics of the scale smaller than $\sim 2.0\ \text{fm}$.

\section{Summary}
\label{sec:summary}
We studied the effects of the hydrodynamic fluctuations during the evolution of the QGP fluids
on rapidity decorrelation.
We employed an integrated dynamical model in which
the hydrodynamic model, {\tt rfh},
implementing the causal hydrodynamic fluctuations and dissipations,
is combined with the hadronic cascade model {\tt JAM}\@.
We performed simulations of Pb+Pb collisions at $\sqrt{s_{NN}}=2.76\ \text{TeV}$
using the integrated dynamical model.
For comparison, we
simulated the hydrodynamic stage with or without the hydrodynamic fluctuations and
with different sets of the cutoff parameters ($\lambda_\perp$ and $\lambda_\eta$).
We fixed the model parameters so that our
model fairly reproduces the experimental data of
the multiplicity normalized by the number of the participant,
the pseudorapidity distributions,
and the transverse momentum dependence of $v_{2}\{2\}$ for charged hadrons.
With this model,
we calculated  the factorization ratio $r_n$ in the longitudinal direction
and its centrality dependence
to estimate the effects of hydrodynamic fluctuations on rapidity decorrelation.
We found that the hydrodynamic fluctuations bring the sizable effects on the factorization ratios:
Due to the nature of hydrodynamic fluctuations being random in space and time,
longitudinal correlations of anisotropic flow tend to break down.
To further understand this,
we calculated the Legendre coefficients of the magnitude and the event-plane angle
of anisotropic flow parameters that characterize
the fluctuations of the longitudinal flow decorrelation.
These Legendre coefficients increase by adding hydrodynamic fluctuations.
We also analyzed the cutoff parameter dependence of the Legendre coefficients
and found that the Legendre coefficients increase with larger $1/\lambda$ (smaller $\lambda$).
These analyses showed that hydrodynamic fluctuations
with a small cutoff parameter below $\lambda \sim 2$ play an important role in
understanding rapidity decorrelation phenomena.

Although we found the significance of the hydrodynamic fluctuations in understanding
longitudinal dynamics in high-energy nuclear collisions,
we did not perfectly reproduce the pseudorapidity
and the centrality dependences of the factorization ratios.
In future, we plan to investigate the effects of initial fluctuations of longitudinal profiles
which are missing in the present study.
In particular, we will quantify how much the initial longitudinal fluctuations
together with the hydrodynamic fluctuations
bring the rapidity decorrelations in describing the factorization ratios.

Since the power of the fluctuating forces given by the FDR depends on temperature explicitly
together with the shear viscous coefficient, the additional temperature dependence appears.
Along the lines of this perspective,
the collision energy dependence of the factorization ratios would be also interesting
since the maximum temperature of the system created in
heavy-ion collisions at the RHIC top energy
would be smaller
compared with the one at LHC.
One could have a chance to
focus more on the dynamics around the transition region ($T\sim 160$ MeV).

\section*{Acknowledgment}
This work was supported by JSPS KAKENHI Grant Numbers JP18J22227 (A.S.) and JP19K21881 (T.H.).
K.M. is supported by the NSFC under Grant No. 11947236.

\appendix
\bibliography{References}

\end{document}